\documentclass[10pt,a4paper]{article}
\usepackage[a4paper, left=2cm, right=2cm, top=1cm]{geometry}

\usepackage[utf8]{inputenc}
\usepackage{amsmath, amsthm, amssymb, amsfonts}     
\usepackage{mathrsfs}                               
\usepackage{accents}                                
\usepackage{natbib}                                 
\usepackage{hyperref}                               
\usepackage{doi}                                    
\usepackage{authblk}                                
\usepackage{siunitx}                                
\usepackage{booktabs}                               
\usepackage{ulem}                                   
\usepackage{nomencl}                                
\makenomenclature                                   
\usepackage[linesnumbered,ruled,vlined]{algorithm2e}
\SetCommentSty{mycommfont}
\usepackage{subfigure}

\usepackage{bm}                     

\usepackage{graphicx}
\usepackage{tikz}
\usepackage{pgfplots}

\usetikzlibrary{patterns}

\usetikzlibrary{decorations.pathmorphing,arrows,intersections,arrows.meta,angles,quotes}

\pgfplotsset{
	kurze Legende/.style={
		legend image code/.code={
			\draw[##1,mark repeat=2,line width=0.6pt]
			plot coordinates {
				(0cm,0cm)
				(0.3cm,0cm)
			};
		}
	}
}

\pgfplotsset{
	compat = newest,
	scale only axis, 
	max space between ticks = 50pt,
	ticklabel style = {font=\footnotesize},
	legend style =  {font=\footnotesize},
	grid = major,
	grid style = {dotted},
	legend columns=1, 
	xtick pos=left,
	ytick pos=left
}

\newcommand{\analytiSolutionPictures}{
	\pgfplotsset{  
		width=0.22\textwidth,
		height=0.275\textwidth,
		ylabel style={text width=0.2\textwidth,align=center}
	}
}

\pgfplotsset{select coords between index/.style 2 args={
		x filter/.code={
			\ifnum\coordindex<#1\fi
			\ifnum\coordindex>#2\fi
		}
}}

\definecolor{color1}{HTML}{0060AD} 
\definecolor{color2}{HTML}{FF4500} 
\definecolor{color3}{HTML}{FFA500} 
\definecolor{color4}{HTML}{006400} 
\definecolor{color5}{HTML}{9400D3} 
\definecolor{color6}{HTML}{800000} 
\definecolor{color7}{HTML}{000000} 
\definecolor{color8}{HTML}{0000FF} 
\definecolor{color9}{HTML}{FF0000} 
\definecolor{mycolor_blue}{RGB}{66,124,161}
\definecolor{mycolor_grey}{RGB}{198,198,198} 

\tikzstyle{line1} = [color=color7,semithick] 
\tikzstyle{line2} = [color=color2,densely dotted,semithick]
\tikzstyle{line3} = [color=color1,densely dashed,semithick]
\tikzstyle{line4} = [color=color5,dash dot,semithick]
\tikzstyle{line5} = [color=color4,dash dot dot,semithick]
\tikzstyle{line6} = [color=color6,semithick]

\tikzstyle{mark1} = [color=color7,mark=x,mark size=2pt,mark options=solid,semithick] 
\tikzstyle{mark2} = [color=color2,mark=o,mark size=2pt,mark options=solid,semithick]
\tikzstyle{mark3} = [color=color1,mark=*,mark size=2pt,mark options=solid,semithick]
\tikzstyle{mark4} = [color=color5,mark=triangle,mark size=2pt,mark options=solid,semithick]
\tikzstyle{mark5} = [color=color4,mark=square,mark size=2pt,mark options=solid,semithick]
\tikzstyle{mark6} = [color=color7,mark=o,mark size=2pt,mark options=solid,semithick]
\tikzstyle{mark7} = [color=color7,mark=*,mark size=2pt,mark options=solid,semithick]
\tikzstyle{mark8} = [color=color7,mark=triangle,mark size=2pt,mark options=solid,semithick]

\usetikzlibrary{external}
\title{On the Removal of Solver-Induced Dependencies in Momentum-Weighted Interpolation for Primal and Continuous-Adjoint Flow Solvers}
\author[]{Niklas K\"uhl\thanks{kuehl@hsva.de}}

\affil[]{Hamburg Ship Model Basin, Bramfelder Strasse 164, D-22305 Hamburg, Germany}

\begin{document}

\providetoggle{tikzExternal}
\settoggle{tikzExternal}{true}
\settoggle{tikzExternal}{false}

\maketitle

\begin{abstract}
Momentum-Weighted Interpolation (MWI) is a key component in pressure--velocity coupling schemes on collocated cell-centered finite-volume methods for both primal and continuous adjoint formulations. In many practical implementations, MWI relies on diagonal momentum coefficients that include contributions from under-relaxation and time discretization. As a result, both primal quantities of interest and adjoint sensitivities may exhibit a non-physical dependence on solver parameters such as relaxation factors and time-step size, and no well-defined limit is obtained as these parameters approach zero.

In this work, building on previous developments in discrete-consistent MWI formulations, a simple correction is proposed that removes solver-induced contributions from the diagonal momentum coefficients in the pressure-driven term. The resulting formulation preserves the original discretization while eliminating artificial dependencies on relaxation and time-stepping parameters and is applied consistently to both primal and adjoint systems. To facilitate its application, the derivation is presented in a structured, recipe-like manner that can be readily followed and transferred to different finite volume-based solver configurations.

The proposed modification is assessed for a two-dimensional laminar cylinder flow and a three-dimensional turbulent ship hull flow configuration. In both cases, the uncorrected formulation leads to significant variations in forces, wake-related quantities, and shape sensitivities when solver parameters are altered, despite all simulations being iterated to converged residual levels and stable integral quantities. In contrast, the corrected formulation yields consistent results across a wide range of relaxation factors and time-step sizes. Deviations between the inconsistent and consistent formulations are found to be on the order of $\mathcal{O}(1\%)$ for force-based quantities, while field-based measures, such as the ship wake considered in this work and generally exhibiting increased sensitivity, may differ by up to $20\%$.

The presented approach provides a simple and effective means to restore solver-independent, physically consistent pressure--velocity coupling for both primal and adjoint formulations with minimal implementation effort.
\end{abstract}

\begin{flushleft}
\small{\textbf{{Keywords:}}} Computational Fluid Dynamics, Momentum-Weighted Interpolation, Rhie--Chow Interpolation, Continuous Adjoint Sensitivity Analysis, Solver-Parameter Dependence
\end{flushleft}


\section{Introduction}


Pressure--velocity coupling in collocated cell-centered finite-volume methods is a critical aspect of Computational Fluid Dynamics (CFD) and requires appropriate flux reconstruction techniques to avoid pressure--velocity decoupling \cite{ferziger2012computational, yu2002checkerboard}. Among these, Momentum-Weighted Interpolation (MWI), closely related to the classical Rhie--Chow procedure \cite{rhie1983numerical, choi2003use}, plays a central role in the computation of face fluxes and thereby directly influences both the stability and consistency of the discrete solution. The consistent treatment of pressure--velocity coupling is essential for both primal and adjoint computations, as inaccuracies in the discrete flux reconstruction affect not only the flow solution itself but also propagate into the adjoint system and influence the resulting sensitivities. MWI can be interpreted as a general framework derived from the semi-discrete momentum equations \cite{pascau2011cell, bartholomew2018unified, mencinger2007finite, yakubov2015experience}.

A discrete-consistent adjoint MWI formulation was introduced in \cite{kuhl2022discrete}, which serves as the foundation for the present work. Building directly upon this formulation, the MWI approach was systematically extended to the adjoint system, with particular emphasis on the treatment of explicitly appearing adjoint source terms, which commonly arise in segregated continuous adjoint solver strategies. It was shown that a careful implementation, including a consistent, reversed levered/weighted interpolation from cell-centered to face-based quantities, is essential to ensure robust convergence of the adjoint solver while maintaining the accuracy of the predicted sensitivities.

However, in practical implementations, the diagonal momentum coefficients entering the MWI formulation typically include contributions from under-relaxation and time discretization. As a consequence, the resulting face fluxes depend not only on the underlying physical discretization but also on solver-specific parameters such as relaxation factors and time step size. This dependence leads to variations in both primal quantities of interest and adjoint sensitivities when such solver parameters are altered, even if the underlying spatial discretization remains unchanged. In particular, no well-defined limit is obtained as the time step approaches zero or relaxation factors are reduced.

Building upon the previously introduced paper in this journal (\cite{kuhl2022discrete}), this study continues the established line of investigation. It addresses the identified issue by introducing a simple modification of the diagonal momentum coefficients used in MWI. By removing contributions from under-relaxation and time discretization prior to interpolation, solver-induced dependencies are eliminated without altering the underlying discretization or solution procedure. The resulting formulation applies consistently to both primal and adjoint systems. In this way, the present work not only extends the findings of the preceding paper but also provides a practical pathway to close the gap between physically consistent flux reconstruction and solver-dependent implementations commonly used in practical CFD solvers.

To ensure clarity and transferability, the derivation is structured into four distinct parts, following a unified, generic procedure. These cover the classical MWI formulation, under-relaxation only, time discretization only, and the combined effect of both. The derivation is presented for the primal system without loss of generality, as both the procedure and the resulting formulation can be transferred directly to the adjoint system.

The derivation is carried out within a standard cell-centered finite-volume framework for unstructured meshes containing arbitrary polyhedral cells. It is based on the semi-discrete momentum balance formulated at both the cell center and, in a formal sense, at the face. By interpolating the cell-centered relation to the faces, a predictor quantity is obtained, which shares non-pressure-driven contributions with the face-based momentum balance. This enables their elimination, providing direct access to the corrected face flux. The different formulations arise from the level of detail with which the momentum balance is expressed, in particular, whether contributions from time discretization and under-relaxation are included in the assembled system or treated separately.

The proposed approach is assessed for a two-dimensional cylinder at low Reynolds number and a three-dimensional ship hull configuration. It is demonstrated that the uncorrected formulation exhibits a pronounced dependence on solver parameters. In contrast, the corrected formulation yields consistent results across a wide range of relaxation factors and time step sizes.

The remainder of this paper is organized as follows. Section~\ref{sec:mwi} briefly reviews the role of MWI and identifies the origin of solver-induced dependencies. Section~\ref{sec:removal} introduces the proposed correction and discusses its implementation. Section~\ref{sec:applications} presents numerical results for the considered test cases. Finally, conclusions are drawn in Section~\ref{sec:conclusion}.

\section{Momentum-Weighted Interpolation and Solver Dependencies}
\label{sec:mwi}

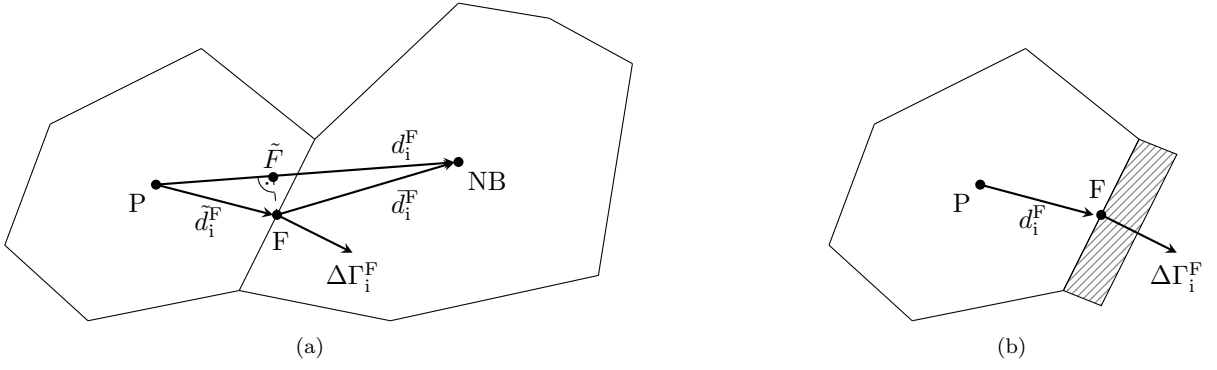
\begin{figure}[!ht]
\centering
\subfigure[]{
\iftoggle{tikzExternal}{
\begin{tikzpicture}

\draw[thin] (0.0,0.0) -- (2.0,0.4);
\draw[thin] (2.0,0.4) -- (3.0,2.4);
\draw[thin] (3.0,2.4) -- (1.5,3.6);
\draw[thin] (1.5,3.6) -- (-0.5,2.6);
\draw[thin] (-0.5,2.6) -- (-1.1,1.0);
\draw[thin] (-1.1,1.0) -- (0.0,0.0);

\node[] at (0.9,1.8) [circle,fill,scale=0.4]{};
\node[anchor=north east] at (0.9,1.8) {P};

\node[] at (2.5,1.4) [circle,fill,scale=0.4]{};
\node[anchor=north] at (2.55,1.3) {F};

\draw[thick,->,>=stealth] (2.5,1.4) -- (3.5,0.9);
\node[anchor=north] at (3.5,0.9) {$\Delta \Gamma_\mathrm{i}^\mathrm{F}$};

\draw[thick,->,>=stealth] (0.9,1.8) -- (4.85,2.095);
\node[anchor=south] at (4.2,2.0) {$d_\mathrm{i}^\mathrm{F}$};

\draw[thick,->,>=stealth] (0.9,1.8) -- (2.45,1.4);
\node[anchor=north] at (1.6,1.65) {$\tilde{d}_\mathrm{i}^\mathrm{F}$};

\draw[thick,->,>=stealth] (2.5,1.4) -- (4.85,2.095);
\node[anchor=north] at (4.2,1.85) {$\bar{d}_\mathrm{i}^\mathrm{F}$};

\draw[thin] (2.0,0.4) -- (4.0,0.0);
\draw[thin] (4.0,0.0) -- (6.75,0.6);
\draw[thin] (6.75,0.6) -- (7.2,3.4);
\draw[thin] (7.2,3.4) -- (6.1,4.0);
\draw[thin] (6.1,4.0) -- (4.9,4.2);
\draw[thin] (4.9,4.2) -- (3.0,2.4);

\node[] at (4.9,2.1) [circle,fill,scale=0.4]{};
\node[anchor=north west] at (4.9,2.1) {NB};

\node[] at (2.45,1.9) [circle,fill,scale=0.4]{};
\node[anchor=south] at (2.45,1.9) {$\tilde{F}$};

\draw[thin, dashed] (2.45,1.9) -- (2.5,1.4);

\draw[] (2.40,1.9) coordinate (P) -- (2.45,1.9) coordinate (Q) -- (2.455,1.85) coordinate (R);
\draw pic [draw, angle radius=2mm, "$\cdot$"] {angle =P--Q--R};

\end{tikzpicture}}{
\includegraphics{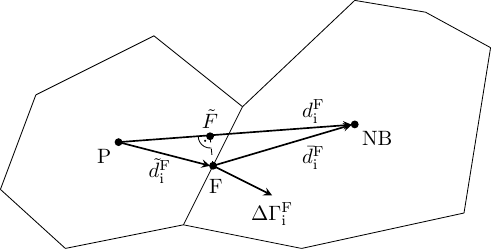}}
}
\hspace{2cm}
\subfigure[]{
\iftoggle{tikzExternal}{
\begin{tikzpicture}

\draw[thin] (0.0,0.0) -- (2.0,0.4);
\draw[thin] (2.0,0.4) -- (3.0,2.4);
\draw[thin] (3.0,2.4) -- (1.5,3.6);
\draw[thin] (1.5,3.6) -- (-0.5,2.6);
\draw[thin] (-0.5,2.6) -- (-1.1,1.0);
\draw[thin] (-1.1,1.0) -- (0.0,0.0);

\filldraw[pattern=north east lines, pattern color=black!50] (2.0,0.4) -- (2.5,0.2) -- (3.5,2.2) -- (3.0,2.4) -- (2.0,0.4);

\node[] at (0.9,1.8) [circle,fill,scale=0.4]{};
\node[anchor=north east] at (0.9,1.8) {P};

\node[] at (2.5,1.4) [circle,fill,scale=0.4]{};
\node[anchor=south] at (2.45,1.5) {F};

\draw[thick,->,>=stealth] (0.9,1.8) -- (2.4,1.4);
\node[anchor=north] at (1.6,1.65) {$d_\mathrm{i}^\mathrm{F}$};

\draw[thick,->,>=stealth] (2.5,1.4) -- (3.5,0.9);
\node[anchor=north] at (3.5,0.9) {$\Delta \Gamma_\mathrm{i}^\mathrm{F}$};


\end{tikzpicture}}{
\includegraphics{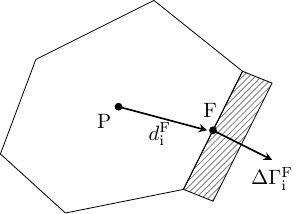}}
}
\caption{Schematic representation of a finite-volume arrangement (a) in the field and (b) at a boundary.}
\label{fig:finite_volume_approximation}
\end{figure}

The derivation of the MWI strategy is presented in the following.
A distinction is made between the cell center $P$ and the set of surrounding faces $F(P)$. All primary variables are stored at the cell centers and must therefore be reconstructed at the faces. A schematic representation of an interior control volume and a boundary control volume is shown in Fig.~\ref{fig:finite_volume_approximation}.

The derivation, therefore, begins with the standard formulation, in which such contributions are already incorporated into the assembled system of equations. For a control volume $P$, the semi-discrete primal linear momentum equation may be written as
\begin{align}
A_\mathrm{P} \, v_{\mathrm{P},\mathrm{i}}
=
H_{\mathrm{P},\mathrm{i}}
-
\Omega_\mathrm{P} \left[
\frac{\partial p}{\partial x_{\mathrm{i}}}- f_{\mathrm{i}}
\right]_\mathrm{P} ,
\label{equ:mome_semi_discrete_std}
\end{align}
where $v_{\mathrm{P},\mathrm{i}}$ denotes the $i$-th component of the primal velocity at the cell center, $A_\mathrm{P}$ the diagonal momentum coefficient, $\Omega_\mathrm{P}$ the control-volume size, $H_{\mathrm{P},\mathrm{i}}$ collects all non-pressure contributions of the discretized momentum balance, and $f_{\mathrm{P},\mathrm{i}}$ represents a volumetric momentum source term.
Since Eqn.~\ref{equ:mome_semi_discrete_std} represents a local balance, it may formally be evaluated not only at cell centers but also at face locations. Solving Eqn.~\ref{equ:mome_semi_discrete_std} for the velocity therefore yields
\begin{align}
v_{\mathrm{P},\mathrm{i}}
=
\frac{H_{\mathrm{P},\mathrm{i}}}{A_\mathrm{P}}
-
\frac{\Omega_\mathrm{P}}{A_\mathrm{P}}
\left[
\frac{\partial p}{\partial x_{\mathrm{i}}} - f_{\mathrm{i}}
\right]_\mathrm{P} 
\qquad \qquad \text{and} \qquad \qquad
v_{\mathrm{F},\mathrm{i}}
=
\frac{H_{\mathrm{P},\mathrm{i}}}{A_\mathrm{P}}\bigg|_\mathrm{F}
-
\frac{\Omega_\mathrm{P}}{A_\mathrm{P}}\bigg|_\mathrm{F}
\left[
\frac{\partial p}{\partial x_{\mathrm{i}}} - f_{\mathrm{i}}
\right]_\mathrm{F}.
\label{equ:up_semi_discrete_std}
\end{align}
At first glance, Eqn.~\ref{equ:up_semi_discrete_std} suggests a staggered arrangement, as the same relation is written at both cell centers and face locations. However, in a collocated finite-volume framework, only cell-centered quantities are available, and face values must be reconstructed from them.
In the following, the focus is on the face-normal volume flux. To this end, both expressions in Eqn.~\ref{equ:up_semi_discrete_std} are projected onto the face area vector $\Delta \Gamma_{\mathrm{F},\mathrm{i}}$. Interpolating the cell-centered expression to the face yields
\begin{align}
\overline{v}_{\mathrm{F},\mathrm{i}}
=
\left(
\frac{H_{\mathrm{P, i}}}{A_\mathrm{P}}
\right)_\mathrm{F}
-
b_\mathrm{F} \,
\overline{\left[\frac{\partial p}{\partial x_{\mathrm{i}}} - f_{\mathrm{i}}\right]}_\mathrm{F},
\end{align}
and thus the corresponding predictor flux
\begin{align}
\widetilde{\dot{V}}_\mathrm{F}
=
\left( \overline{v}_{\mathrm{i}} \, \Delta  \Gamma_{\mathrm{i}} \right)_\mathrm{F}
=
\left(
\frac{H_{\mathrm{P,i}}}{A_\mathrm{P}}
\right)_\mathrm{F} \Delta \Gamma_{\mathrm{F},\mathrm{i}}
-
b_\mathrm{F} \,
\overline{\left[\frac{\partial p}{\partial x_{\mathrm{i}}} - f_{\mathrm{i}}\right]}_\mathrm{F} \Delta \Gamma_{\mathrm{F},\mathrm{i}} \, .
\label{equ:mwi_predictor}
\end{align}
Here, $b_\mathrm{F}$ denotes a face-based mobility coefficient resulting from the interpolation of $\Omega_\mathrm{P}/A_\mathrm{P}$ to the face.
The predictor flux is readily available, as it follows directly from the interpolated cell-centered velocity field. However, it contains the non-pressure-driven contribution $\left(H_{\mathrm{P, i}}/A_\mathrm{P}\right)_\mathrm{F}$, which can be eliminated by projecting the face-based form of Eqn.~\ref{equ:up_semi_discrete_std} onto the face-normal direction, yielding
\begin{align}
\dot{V}_\mathrm{F}
=
\frac{H_{\mathrm{P},\mathrm{i}}}{A_\mathrm{P}}\bigg|_\mathrm{F} \Delta \Gamma_{\mathrm{F},\mathrm{i}}
-
b_\mathrm{F}
\left[
\frac{\partial p}{\partial x_{\mathrm{i}}}- f_{\mathrm{i}}
\right]_\mathrm{F} \Delta \Gamma_{\mathrm{F},\mathrm{i}}.
\label{equ:mwi_face_balance}
\end{align}
Equations~\ref{equ:mwi_predictor} and \ref{equ:mwi_face_balance} contain the same non-pressure-driven contribution. Eliminating this term yields a correction to the predictor flux and results in the classical MWI formulation
\begin{align}
\dot{V}_\mathrm{F}
=
\widetilde{\dot{V}}_\mathrm{F}
-
b_\mathrm{F}
\left[
\left[\frac{\partial p}{\partial x_{\mathrm{i}}}- f_{\mathrm{i}}\right]_\mathrm{F} 
-
\overline{\left[\frac{\partial p}{\partial x_{\mathrm{i}}} - f_{\mathrm{i}}\right]}_\mathrm{F}
\right] \Delta \Gamma_{\mathrm{F},\mathrm{i}}.
\label{equ:mwi_final}
\end{align}
Equation~\ref{equ:mwi_final} corresponds to the classical MWI formulation including external momentum sources. It shows that the final face flux is obtained by augmenting the predictor flux with a correction term that replaces the interpolated gradient contribution by a face-based evaluation consistent with the discrete momentum balance, weighted by the mobility coefficient $b_\mathrm{F}$, which thus represents the MWI name-giving weighting factor.

Here, the overbar, e.g., $\overline{\phi}_\mathrm{F}$, denotes a face value obtained by interpolation of cell-centered quantities, typically using linear interpolation. In contrast, the non-overlined quantities $\phi_\mathrm{F}$ denote face-based evaluations that are consistent with the discrete momentum balance.
In practical finite-volume implementations, both terms are reconstructed from cell-centered data. The interpolated quantity $\overline{\left[\frac{\partial p}{\partial x_{\mathrm{i}}} - f_{\mathrm{i}}\right]}_\mathrm{F}$ is obtained by applying a standard interpolation procedure, whereas the face-based quantity $\left[\frac{\partial p}{\partial x_{\mathrm{i}}}\big|_\mathrm{F} - f_{\mathrm{F},\mathrm{i}}\right]$ is evaluated using a discretization consistent with the flux formulation of the momentum equations, e.g., based on a face-normal difference (CDS-type evaluation).

It is important to recognize that both expressions originate from the same underlying continuous fields and are therefore equivalent in the continuous limit. However, at the discrete level, they differ due to distinct reconstruction operators. This discrepancy remains finite on practical grids and is precisely what suppresses spurious pressure--velocity decoupling.

However, the mobility coefficient $b_\mathrm{F}$ scales with the inverse diagonal momentum coefficient, i.e.,
\begin{align}
b_\mathrm{F} \sim \frac{\Omega_\mathrm{P, F}^{\mathrm{proj}}}{A_{\mathrm{P},\mathrm{F}}}.
\end{align}
Here, $\Omega_\mathrm{P, F}^{\mathrm{proj}} = [d_\mathrm{i} \, \Delta \Gamma_\mathrm{i}]_\mathrm{F}$ denotes the projected control-volume contribution associated with the face area vector, cf.~Fig.~\ref{fig:finite_volume_approximation}, while the diagonal momentum coefficient $A_{\mathrm{P},\mathrm{F}}$ is obtained by linear interpolation of the neighboring cell-centered values. Consequently, contributions to $A_{\mathrm{P},\mathrm{F}}$ arising from relaxation or time discretization directly affect the magnitude of $b_\mathrm{F}$. In particular, these contributions usually scale with $1/\Delta t$ and $1/\omega$ (cf. \cite{ferziger2012computational}, such that an increase in diagonal dominance leads to a reduction of $b_\mathrm{F}$. This, in turn, weakens the pressure-correction term and degrades the quality of the flux correction.

This effect is particularly relevant for adjoint formulations, where smaller (pseudo-)time-step sizes or relaxation factors are typically required due to the increased stiffness of the governing equations arising from additional explicit coupling terms, cf. \cite{kuhl2021continuous, kuhl2022discrete, bletsos2023adjoint}. As a result, the degradation of the MWI correction may be even more pronounced in the adjoint system. This behavior motivates the modifications introduced in the following.

\section{Removal of Solver-Induced Contributions}
\label{sec:removal}
The formulation presented above is extended to a more general case, in which both time discretization and under-relaxation of the momentum equations are present and consistently removed from the MWI correction. This combined treatment reflects the typical structure of practical segregated primal and adjoint flow solvers and allows for a consistent assessment of their influence on the flux reconstruction.
For completeness, the individual effects of relaxation and time discretization are discussed separately in Appendices~\ref{sec:derivation_mwi_relax} and \ref{sec:derivation_mwi_time}, where the corresponding intermediate formulations are derived and analyzed.

In practical unsteady segregated solution procedures, the momentum equations include additional contributions from time discretization, which are subsequently solved in an under-relaxed form. Here, $n$ denotes the current time level and $k$ the current outer, i.e., pressure-velocity coupling, iteration within that time level. Considering implicit Euler time integration, the semi-discrete momentum Eqn. \ref{equ:mome_semi_discrete_std} may be written as
\begin{align}
\frac{1}{\omega}
\left(
A_\mathrm{P} + \frac{\rho_\mathrm{P} \Omega_\mathrm{P}}{\Delta t}
\right)^\mathrm{n,k} v_{\mathrm{P},\mathrm{i}}^\mathrm{n,k}
&=
H_{\mathrm{P},\mathrm{i}}^\mathrm{n,k}
-
\Omega_\mathrm{P}\left[
\frac{\partial p}{\partial x_{\mathrm{i}}} - f_{\mathrm{i}}
\right]_\mathrm{P}^\mathrm{n,k}
+
\frac{\rho_\mathrm{P} \Omega_\mathrm{P}}{\Delta t}\,v_{\mathrm{P},\mathrm{i}}^\mathrm{n-1}
+
\frac{1-\omega}{\omega}
\left(
A_\mathrm{P} + \frac{\rho_\mathrm{P} \Omega_\mathrm{P}}{\Delta t}
\right)^\mathrm{n,k-1} v_{\mathrm{P},\mathrm{i}}^\mathrm{n,k},
\label{equ:time_relax_momentum}
\end{align}
where $\Delta t$ denotes the fixed time-step size, $\rho_\mathrm{P}$ the density, and $\omega \in (0,1]$ the momentum relaxation factor. Here, $A_\mathrm{P}$ denotes the diagonal momentum coefficient arising exclusively from the spatial discretization of the momentum equations, i.e., from convective, diffusive, and other spatially discretized implicit contributions, and is therefore treated independently of the temporal and iterative contributions introduced separately in Eqn.~\ref{equ:time_relax_momentum}.
Dividing Eqn.~\ref{equ:time_relax_momentum} by $A_\mathrm{P}$ yields
\begin{align}
\frac{1}{\omega}
\left(
1 + d_\mathrm{P}^\mathrm{n}
\right)v_{\mathrm{P},\mathrm{i}}^\mathrm{n,k}
&=
\frac{H_{\mathrm{P},\mathrm{i}}^\mathrm{n,k}}{A_\mathrm{P}}
-
\frac{\Omega_\mathrm{P}}{A_\mathrm{P}}
\left[
\frac{\partial p}{\partial x_{\mathrm{i}}} - f_{\mathrm{i}}
\right]_\mathrm{P}^\mathrm{n,k}
+
d_\mathrm{P}^\mathrm{n}\,v_{\mathrm{P},\mathrm{i}}^\mathrm{n-1}
+
\frac{1-\omega}{\omega}
\left(
1 + d_\mathrm{P}^\mathrm{n}
\right)v_{\mathrm{P},\mathrm{i}}^\mathrm{n,k-1},
\label{equ:time_relax_scaled}
\end{align}
with the cell-based temporal scaling factor
\begin{align}
d_\mathrm{P}^\mathrm{n}
=
\frac{\rho_\mathrm{P} \Omega_\mathrm{P}}{A_\mathrm{P} \Delta t} \bigg|^\mathrm{n}.
\end{align}
Multiplying Eqn.~\ref{equ:time_relax_scaled} by $\omega$ gives
\begin{align}
\left(
1 + d_\mathrm{P}^\mathrm{n}
\right)v_{\mathrm{P},\mathrm{i}}^\mathrm{n,k}
&=
\omega
\left[
\frac{H_{\mathrm{P},\mathrm{i}}}{A_\mathrm{P}}
-
\frac{\Omega_\mathrm{P}}{A_\mathrm{P}}
\left(
\frac{\partial p}{\partial x_{\mathrm{i}}} - f_{\mathrm{i}}
\right)_\mathrm{P}
\right]^\mathrm{n,k}
+
\omega \, d_\mathrm{P}^\mathrm{n}\,v_{\mathrm{P},\mathrm{i}}^\mathrm{n-1}
+
(1-\omega)
\left(
1 + d_\mathrm{P}^\mathrm{n}
\right)v_{\mathrm{P},\mathrm{i}}^\mathrm{n,k-1}.
\label{equ:up_time_relax}
\end{align}
The above expression is interpreted both at the cell center and, formally, at the face. Interpolating the cell-centered expression to face $F$ and projecting onto the face area vector yields the predictor flux
\begin{align}
\left(
1 + d_\mathrm{F}^\mathrm{n}
\right)\widetilde{\dot{V}}_\mathrm{F}^\mathrm{n,k}
&=
\omega
\left(
\frac{H_{\mathrm{P,i}}}{A_\mathrm{P}}
\right)_\mathrm{F}^\mathrm{n,k} \Delta \Gamma_{\mathrm{F},\mathrm{i}}
-
\omega \, b_\mathrm{F}^\mathrm{n}
\overline{\left[\frac{\partial p}{\partial x_{\mathrm{i}}} - f_{\mathrm{i}}\right]}_\mathrm{F}^\mathrm{n,k} \Delta \Gamma_{\mathrm{F},\mathrm{i}}
+
\omega \, d_\mathrm{F}^\mathrm{n}\,\widetilde{\dot{V}}_\mathrm{F}^\mathrm{n-1}
+
(1-\omega)
\left(
1 + d_\mathrm{F}^\mathrm{n}
\right)\widetilde{\dot{V}}_\mathrm{F}^\mathrm{n,k-1},
\label{equ:mwi_predictor_time_relax}
\end{align}
where $d_\mathrm{F}^\mathrm{n}$ denotes the face-based temporal scaling factor obtained from interpolation of $d_\mathrm{P}^\mathrm{n}$ to the face.
In complete analogy to Eqn.~\ref{equ:mwi_face_balance_relax}, a face-based form of Eqn.~\ref{equ:up_time_relax} can be written as
\begin{align}
\left(
1 + d_\mathrm{F}^\mathrm{n}
\right)\dot{V}_\mathrm{F}^\mathrm{n,k}
&=
\omega
\left(
\frac{H_{\mathrm{P,i}}}{A_\mathrm{P}}
\right)_\mathrm{F}^\mathrm{n,k} \Delta \Gamma_{\mathrm{F},\mathrm{i}}
-
\omega \, b_\mathrm{F}^\mathrm{n}
\left[
\frac{\partial p}{\partial x_{\mathrm{i}}}- f_{\mathrm{i}}
\right]_\mathrm{F}^\mathrm{n,k}  \Delta \Gamma_{\mathrm{F},\mathrm{i}}
+
\omega \, d_\mathrm{F}^\mathrm{n}\,\dot{V}_\mathrm{F}^\mathrm{n-1}
+
(1-\omega)
\left(
1 + d_\mathrm{F}^\mathrm{n}
\right)\dot{V}_\mathrm{F}^\mathrm{n,k-1},
\label{equ:mwi_face_balance_time_relax}
\end{align}
Equations~\ref{equ:mwi_predictor_time_relax} and \ref{equ:mwi_face_balance_time_relax} again contain the same non-pressure-driven contribution. Eliminating this term yields
\begin{align}
\left(
1 + d_\mathrm{F}^\mathrm{n}
\right)\dot{V}_\mathrm{F}^\mathrm{n,k}
&=
\left(
1 + d_\mathrm{F}^\mathrm{n}
\right)\widetilde{\dot{V}}_\mathrm{F}^\mathrm{n,k}
-
\omega \, b_\mathrm{F}^\mathrm{n}
\left[
\left[\frac{\partial p}{\partial x_{\mathrm{i}}}- f_{\mathrm{i}}\right]_\mathrm{F}^\mathrm{n,k} 
-
\overline{\left[\frac{\partial p}{\partial x_{\mathrm{i}}} - f_{\mathrm{i}}\right]}_\mathrm{F}^\mathrm{n,k}
\right] \Delta \Gamma_{\mathrm{F},\mathrm{i}}
\nonumber\\
&\quad
+
\omega \, d_\mathrm{F}^\mathrm{n}
\left(
\dot{V}_\mathrm{F}^\mathrm{n,k-1}
-
\widetilde{\dot{V}}_\mathrm{F}^\mathrm{n,k-1}
\right)
+
(1-\omega)
\left(
1 + d_\mathrm{F}^\mathrm{n}
\right)
\left(
\dot{V}_\mathrm{F}^\mathrm{n,k-1}
-
\widetilde{\dot{V}}_\mathrm{F}^\mathrm{n,k-1}
\right)
\label{equ:mwi_time_relax_intermediate}
\end{align}
Introducing the stored MWI corrections from the previous time level and the previous outer iteration,
\begin{align}
\Delta \dot{V}_\mathrm{F}^\mathrm{n-1}
=
\dot{V}_\mathrm{F}^\mathrm{n-1}
-
\widetilde{\dot{V}}_\mathrm{F}^\mathrm{n-1}
\qquad \qquad \text{and} \qquad \qquad 
\Delta \dot{V}_\mathrm{F}^\mathrm{n,k-1}
=
\dot{V}_\mathrm{F}^\mathrm{n,k-1}
-
\widetilde{\dot{V}}_\mathrm{F}^\mathrm{n,k-1},
\end{align}
the final time-dependent and under-relaxed MWI formulation can be written in compact recursive form as
%
\begin{align}
\dot{V}_\mathrm{F}^\mathrm{n,k}
&=
\widetilde{\dot{V}}_\mathrm{F}^\mathrm{n,k}
-
\frac{\omega \, b_\mathrm{F}^\mathrm{n}}{1+d_\mathrm{F}^\mathrm{n}}
\left[
\left[\frac{\partial p}{\partial x_{\mathrm{i}}} - f_\mathrm{i}\right]_\mathrm{F}^\mathrm{n,k}
-
\overline{\left[\frac{\partial p}{\partial x_{\mathrm{i}}} - f_{\mathrm{i}}\right]}_\mathrm{F}^\mathrm{n,k}
\right] \Delta \Gamma_{\mathrm{F},\mathrm{i}}
+
\frac{\omega \, d_\mathrm{F}^\mathrm{n}}{1+d_\mathrm{F}^\mathrm{n}}\,\Delta \dot{V}_\mathrm{F}^\mathrm{n-1}
+
(1-\omega)\,\Delta \dot{V}_\mathrm{F}^\mathrm{n,k-1}.
\label{equ:mwi_time_relax_final_div}
\end{align}
The present reformulation removes both temporal and relaxation-induced contributions from the mobility coefficient entering the MWI correction, while retaining the correct transient and iterative updates through the recursive terms proportional to $\Delta \dot{V}_\mathrm{F}^\mathrm{n-1}$ and $\Delta \dot{V}_\mathrm{F}^\mathrm{n,k-1}$. As a result, the pressure-correction scaling becomes independent of both the chosen time-step size and the relaxation parameter. In contrast, the latter affects only the weighting of the current and previously stored corrections.

The formulation can be straightforwardly extended to alternative time-integration schemes, such as implicit three-time-level (ITTL) methods, requiring only a consistent modification of the baseline relation, cf.~Eqn.~\ref{equ:time_relax_momentum}.

\section{Numerical Results}
\label{sec:applications}
The above-described primal and adjoint MWI formulations are implemented in the finite-volume solver FreSCo+. All results are compared relative to different solver configurations, and no attempt is made to validate the solver against reference data. The underlying numerical code has been extensively validated, cf.~\cite{xing2015resistance, lucke2017efd, kroger2018adjoint, angerbauer2020hybrid, andersson2022ship, kuhl2021cahn}. Spatial approximation is of second-order accuracy, while temporal integration is based on a first-order implicit Euler scheme. Convective primal fluxes are approximated using a limited QUICK scheme, whereas diffusive terms are treated with a central differencing approach. In adjoint mode, convective fluxes are approximated using the corresponding QDICK scheme, while diffusive terms are again treated with central differencing, preserving the self-adjoint character of the approximation. For turbulent flows, the solver is operated in RANS mode using the 2003 version of the Menter SST turbulence model, cf. \cite{menter2003ten}. High-Reynolds wall functions are employed for near-wall treatment.

\subsection{Two-Dimensional Laminar Cylinder Flow}
To assess the influence of solver-induced contributions on the MWI, a two-dimensional laminar flow around a circular cylinder is considered. The flow is characterized by a Reynolds number $\mathrm{Re} = \rho V (2R) / \mu = 10$, where $\rho$ denotes the fluid density, $V$ the free-stream velocity, $R$ the cylinder radius, and $\mu$ the dynamic viscosity. The Reynolds number is chosen such that the flow remains steady and laminar.

\begin{figure}[!ht]
\centering
\subfigure[]{
\includegraphics[height=0.3\textwidth]{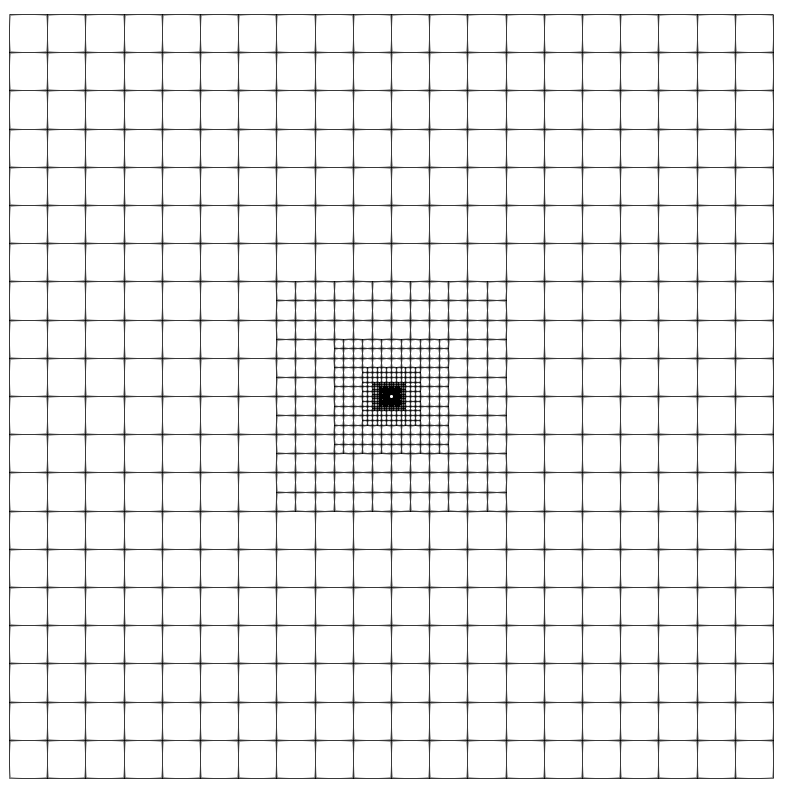}
}
\subfigure[]{
\includegraphics[height=0.3\textwidth]{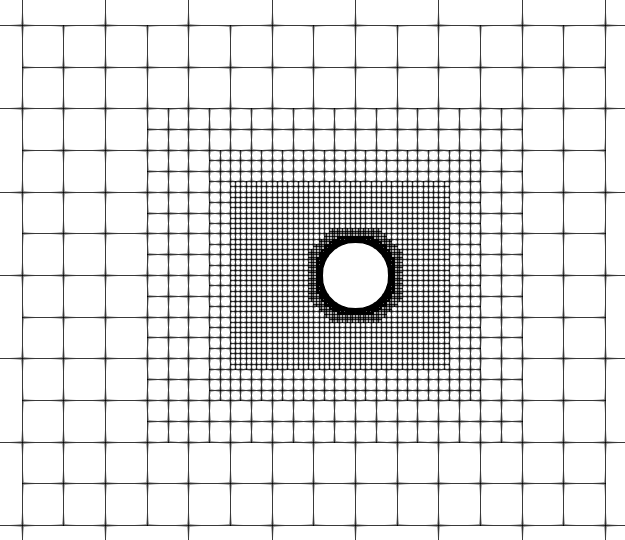}
}
\caption{Submerged cylinder case ($\mathrm{Re}_\mathrm{D} = 10$): Overview of the computational grid (a) and near-wall resolution (b).}
\label{fig:cylinder_grid}
\end{figure}

The computational domain consists of a circular cylinder of radius $R$ placed in a rectangular channel of dimensions $400\,R \times 400\,R$. A uniform inflow velocity is prescribed at the inlet, while a pressure condition is applied at the outlet. No-slip boundary conditions are imposed on the cylinder surface, and symmetry (or slip) conditions are used at the lateral boundaries. The flow regime is chosen such that a steady laminar solution is obtained. An impression of the employed unstructured computational grid that consists of approximately \SI{12000}{} control volumes is shown in Fig.~\ref{fig:cylinder_grid}.

Two groups of numerical experiments are conducted in order to isolate the influence of solver parameters:
\begin{itemize}
    \item \textbf{E1.1:} Steady simulations with varying momentum under-relaxation. The relaxation factor is varied between $6\times10^{-2} \leq \omega < 1$. Results obtained with the classical MWI formulation from Eqn.~\ref{equ:mwi_final} are considered.
    \item \textbf{E1.2:} Steady simulations under identical conditions using the proposed formulation, i.e., Eqn.~\ref{equ:mwi_relax_final}.
    \item \textbf{E2.1:} Pseudo-unsteady simulations with implicit time integration using the classical MWI formulation from Eqn.~\ref{equ:mwi_final}, in which neither temporal nor relaxation contributions are excluded in the mobility coefficient. The normalized time-step size is varied over the range $10^{-2} \leq \Delta t / (V/D) \leq 1$. The momentum relaxation factor for all pseudo-unsteady experiments is $\omega = 0.6$.
    \item \textbf{E2.2:} Pseudo-unsteady simulations excluding temporal contributions in the MWI formulation, cf.~Eqn.~\ref{equ:mwi_time_final}, with identical variation of the time-step size.
    \item \textbf{E2.3:} Pseudo-unsteady simulations excluding relaxation effects only, cf.~Eqn.~\ref{equ:mwi_relax_final}, again using varying time-step sizes.
    \item \textbf{E2.4:} Pseudo-unsteady simulations excluding both temporal and relaxation contributions, cf.~Eqn.~\ref{equ:mwi_time_relax_final_div}.
\end{itemize}
The primary quantities of interest are the primal drag coefficient, i.e., $D / (2 \rho V^2 R^2)$, and the corresponding normalized drag sensitivity derivative, $S^\mathrm{D} / (\rho V^2 R)$, obtained from the adjoint system. Here, $D$ denotes the drag force acting in the streamwise direction, and $S^\mathrm{D}$ represents the corresponding shape sensitivity with respect to the drag, viz.
\begin{align}
D = \int \left[ p \delta_\mathrm{ik} - \mu \left( \frac{\partial v_\mathrm{i}}{\partial x_\mathrm{k}} + \frac{\partial v_\mathrm{k}}{\partial x_\mathrm{i}} \right) \right] n_\mathrm{k} \delta_{i1} \, \mathrm{d} \Gamma \, ,
\qquad \qquad \text{and} \qquad \qquad
S^\mathrm{D} = - \int \mu \, n_\mathrm{m} \frac{\partial v_\mathrm{i}}{\partial x_m} \frac{\partial \hat{v}_\mathrm{i}}{\partial x_l} n_\mathrm{l} \, \mathrm{d} \Gamma \, , \label{equ:cylinder_qoi}
\end{align}
where $n_m$ denotes the entries of the local surface normal vector. In the above expressions, $p$ denotes the primal pressure field, $v_\mathrm{i}$ the primal velocity field, and $\hat{v}_\mathrm{i}$ the adjoint velocity field. The drag force is obtained by integrating the primal fluid stresses, including both pressure and viscous contributions, over the cylinder surface and projecting them onto the inflow direction. The shape sensitivity is derived from the inner product of the gradients of the primal and adjoint velocity fields. The derivation of the integrated shape sensitivity is omitted here for brevity; the reader is referred to \cite{othmer2008continuous, kuhl2021adjoint_2}.

All steady primal and adjoint simulations are iterated close to machine precision and terminated once the solver residuals fall below $10^{-10}$, cf.~\cite{kuhl2022discrete}. Pseudo-unsteady simulations are performed over multiple equivalent flow passages until no further variations in the integral force quantities are observed. In both the steady and pseudo-unsteady cases, the computational effort required to reach convergence increases significantly for smaller time-step sizes and momentum relaxation factors.
All adjoint computations are performed based on a fixed primal solution obtained with the solver-independent formulation, i.e., E1.2 for the steady case and E2.4 for the unsteady case, thereby eliminating any influence of solver-induced effects on the reference state.

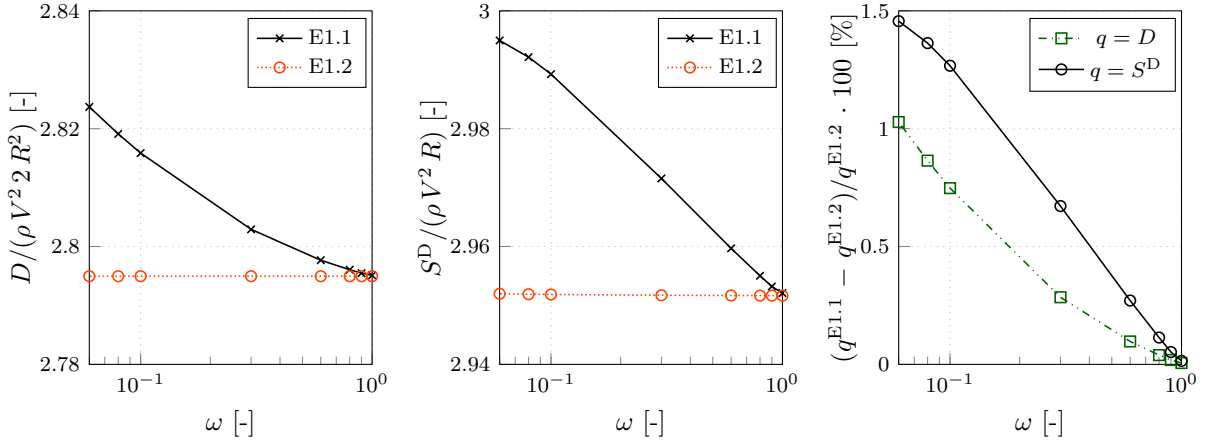
\begin{figure}[!ht]
\centering
\iftoggle{tikzExternal}{
\analytiSolutionPictures
\begin{tikzpicture}
\begin{axis}[
 xlabel={$\omega$ [-]},
 xlabel style={text width=0.25\textwidth,align=center},
 ylabel={$D/(\rho \, V^2 \, 2 \, R^2)$ [-]},
 ylabel shift = -2mm,
 ylabel style={text width=0.35\textwidth,align=center},
 legend style={at={(0.98,0.98)},anchor=north east},
 xmode=log,
 xmin=6E-02,
 xmax=1E-00,
 ymin=2.78,
 ymax=2.84,
]

\addplot [line1, mark1, each nth point=1] table[x expr={\thisrowno{0}},y expr={\thisrowno{1}}]{data/cylinder_results_steady.dat};
\addplot [line2, mark2, each nth point=1] table[x expr={\thisrowno{0}},y expr={\thisrowno{2}}]{data/cylinder_results_steady.dat};

\addlegendentry{E1.1};
\addlegendentry{E1.2};

\end{axis}
\end{tikzpicture}
\begin{tikzpicture}
\begin{axis}[
 xlabel={$\omega$ [-]},
 xlabel style={text width=0.25\textwidth,align=center},
 ylabel={$S^\mathrm{D}/(\rho \, V^2 \, R)$ [-]},
 ylabel shift = -2mm,
 ylabel style={text width=0.35\textwidth,align=center},
 legend style={at={(0.98,0.98)},anchor=north east},
 xmode=log,
 xmin=6E-02,
 xmax=1E-00,
 ymin=2.94,
 ymax=3.00,
]

\addplot [line1, mark1, each nth point=1] table[x expr={\thisrowno{0}},y expr={\thisrowno{3}}]{data/cylinder_results_steady.dat};
\addplot [line2, mark2, each nth point=1] table[x expr={\thisrowno{0}},y expr={\thisrowno{4}}]{data/cylinder_results_steady.dat};

\addlegendentry{E1.1};
\addlegendentry{E1.2};

\end{axis}
\end{tikzpicture}
\begin{tikzpicture}
\begin{axis}[
 xlabel={$\omega$ [-]},
 xlabel style={text width=0.25\textwidth,align=center},
 ylabel={$(q^\mathrm{E1.1} - q^\mathrm{E1.2})/q^\mathrm{E1.2} \cdot 100$ [\%]},
 ylabel shift = -2mm,
 ylabel style={text width=0.35\textwidth,align=center},
 legend style={at={(0.98,0.98)},anchor=north east},
 xmode=log,
 xmin=6E-02,
 xmax=1E-00,
 ymin=0.0,
 ymax=1.5,
]

\addplot [line5, mark5, each nth point=1] table[x expr={\thisrowno{0}},y expr={ ((\thisrowno{1}-\thisrowno{2})/\thisrowno{2})*100 }]{data/cylinder_results_steady.dat};
\addplot [line6, mark6, each nth point=1] table[x expr={\thisrowno{0}},y expr={ ((\thisrowno{3}-\thisrowno{4})/\thisrowno{4})*100 }]{data/cylinder_results_steady.dat};

\addlegendentry{$q = D$};
\addlegendentry{$q = S^\mathrm{D}$};

\end{axis}
\end{tikzpicture}
}{
\includegraphics{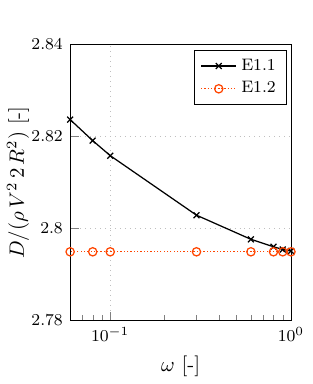}
\includegraphics{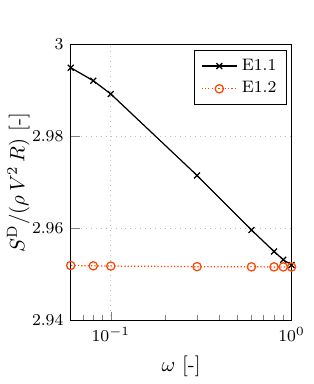}
\includegraphics{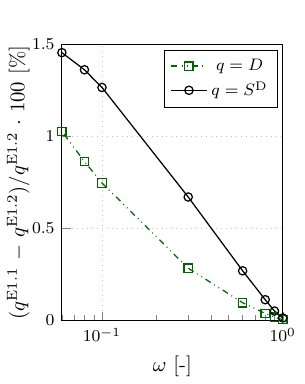}
}
\caption{Submerged cylinder case ($\mathrm{Re}_\mathrm{D} = 10$): Influence of the relaxation factor $\omega$ on the primal drag coefficient (left) and the corresponding adjoint sensitivity (center) for Experiments E1.1 and E1.2. The right-hand side shows the relative deviation of both quantities.}
\label{fig:cylinder_steady_results}
\end{figure}

The corresponding results for the steady simulations are shown in Fig.~\ref{fig:cylinder_steady_results}, where the primal drag coefficient (left) and the corresponding adjoint sensitivity (center) are plotted over the relaxation factor $\omega$ for the classical (E1.1) and the proposed relaxation-free steady MWI formulation (E1.2). It is clearly observed that the predictions obtained with the classical formulation (E1.1) are not constant with respect to the relaxation parameter, but increasingly deviate from the constant results of the proposed formulation (E1.2) as the relaxation factor is reduced. In the limit $\omega \to 1$, both formulations yield identical results, as expected. The increasing discrepancy is also reflected in the relative deviation shown on the right, where the relative difference $(q^{\mathrm{E1.1}} - q^{\mathrm{E1.2}})/q^{\mathrm{E1.2}} \cdot 100$ [\%] is reported. The deviations in the adjoint sensitivity reach up to approximately $1.5\%$, exceeding those of the primal drag coefficient, which are on the order of $1\%$.

\begin{figure}[!ht]
\centering
\subfigure[]{
\includegraphics[width=0.20\textwidth, trim=10cm 3cm 10cm 3cm, clip]{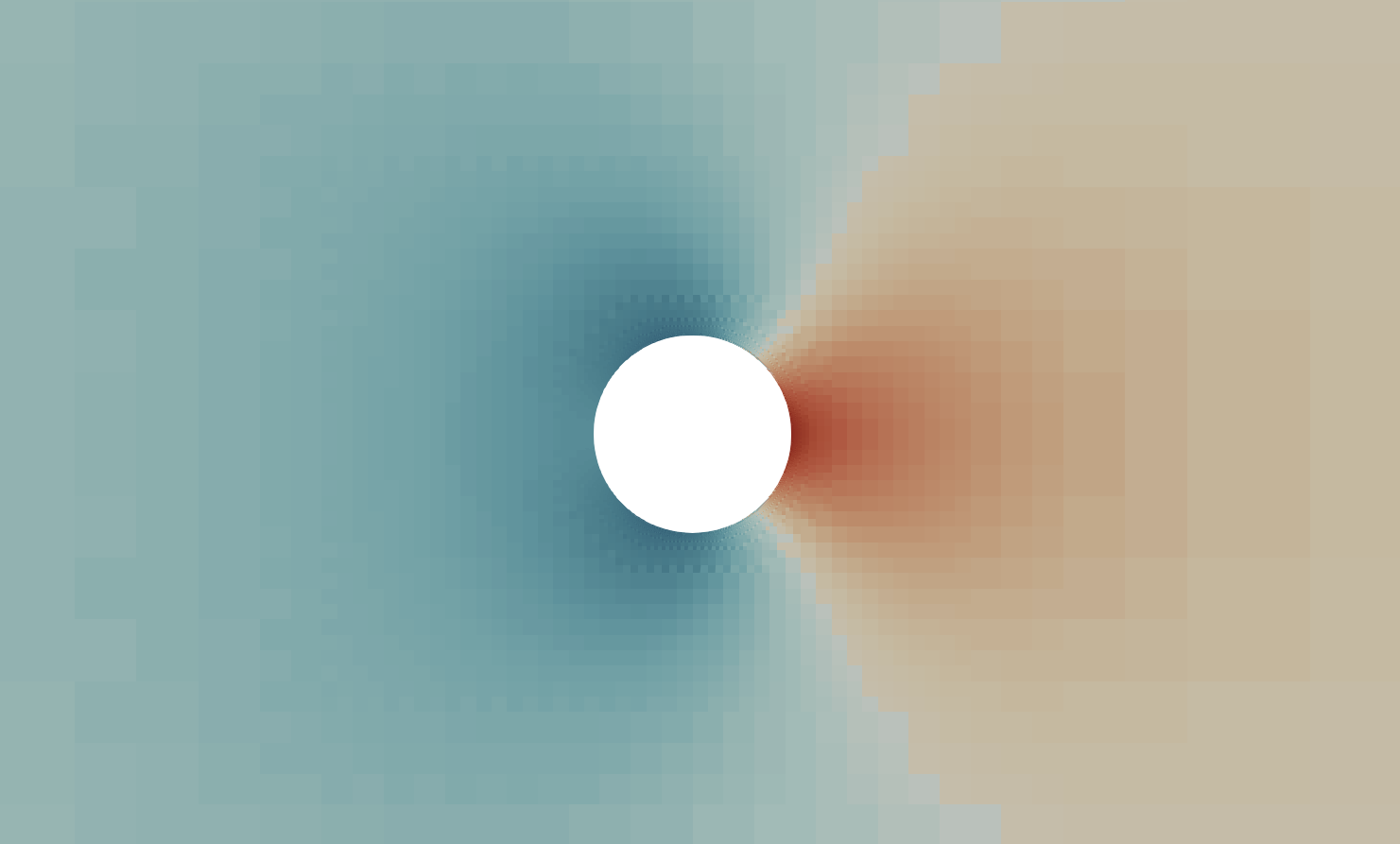}
}
\subfigure[]{
\includegraphics[width=0.20\textwidth, trim=10cm 3cm 10cm 3cm, clip]{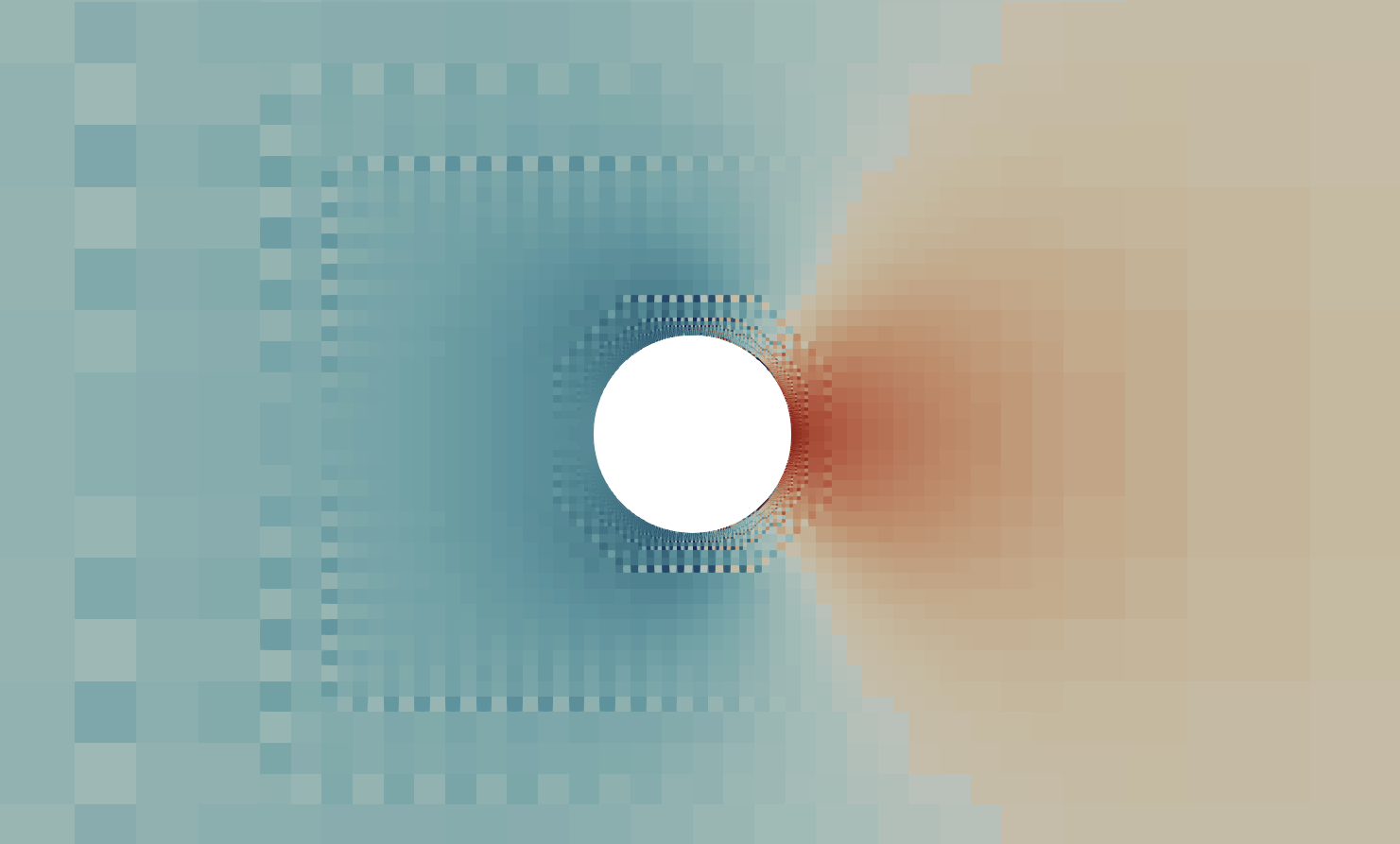}
}
\subfigure[]{
\includegraphics[width=0.20\textwidth, trim=10cm 3cm 10cm 3cm, clip]{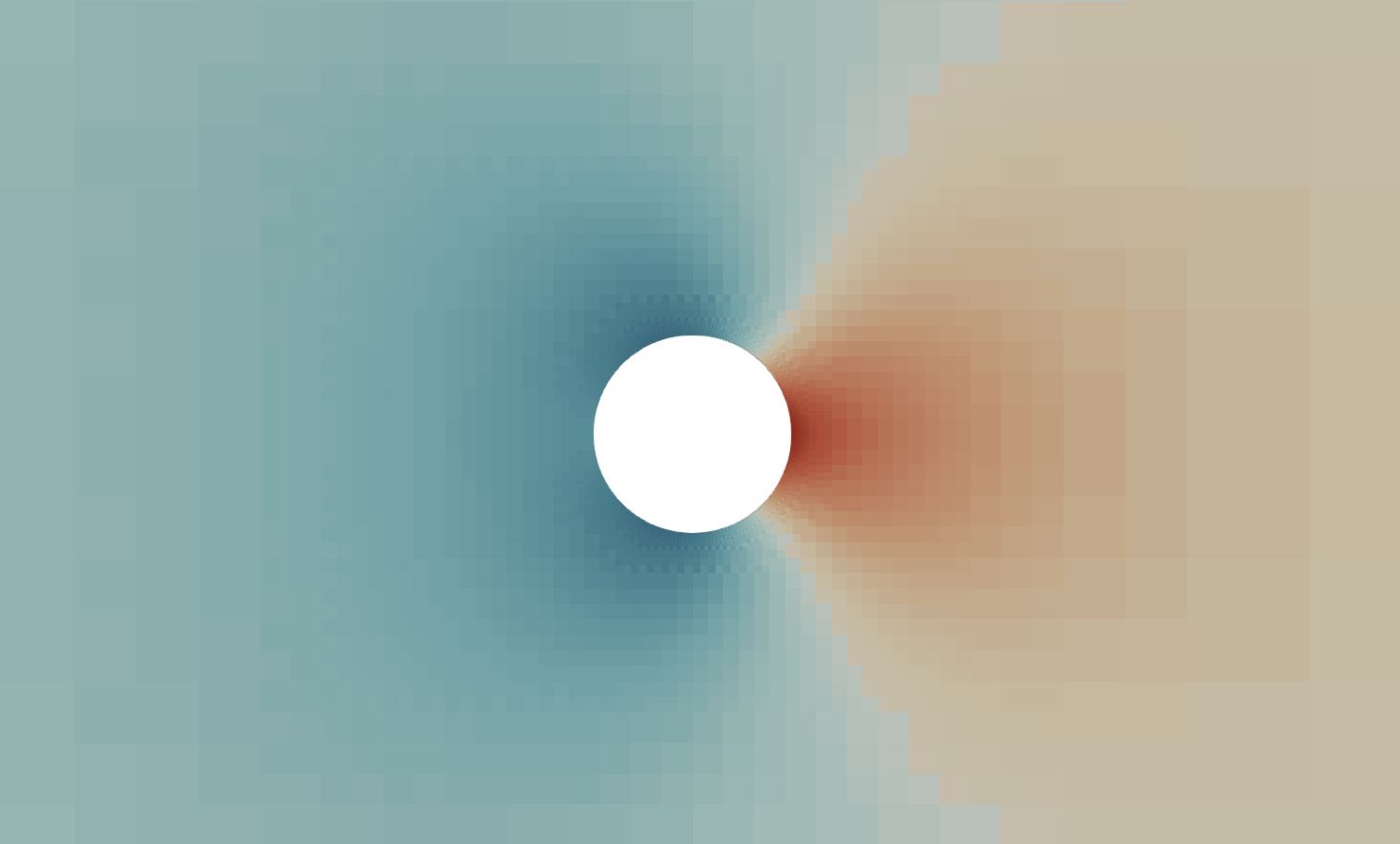}
}
\subfigure[]{
\includegraphics[width=0.20\textwidth, trim=10cm 3cm 10cm 3cm, clip]{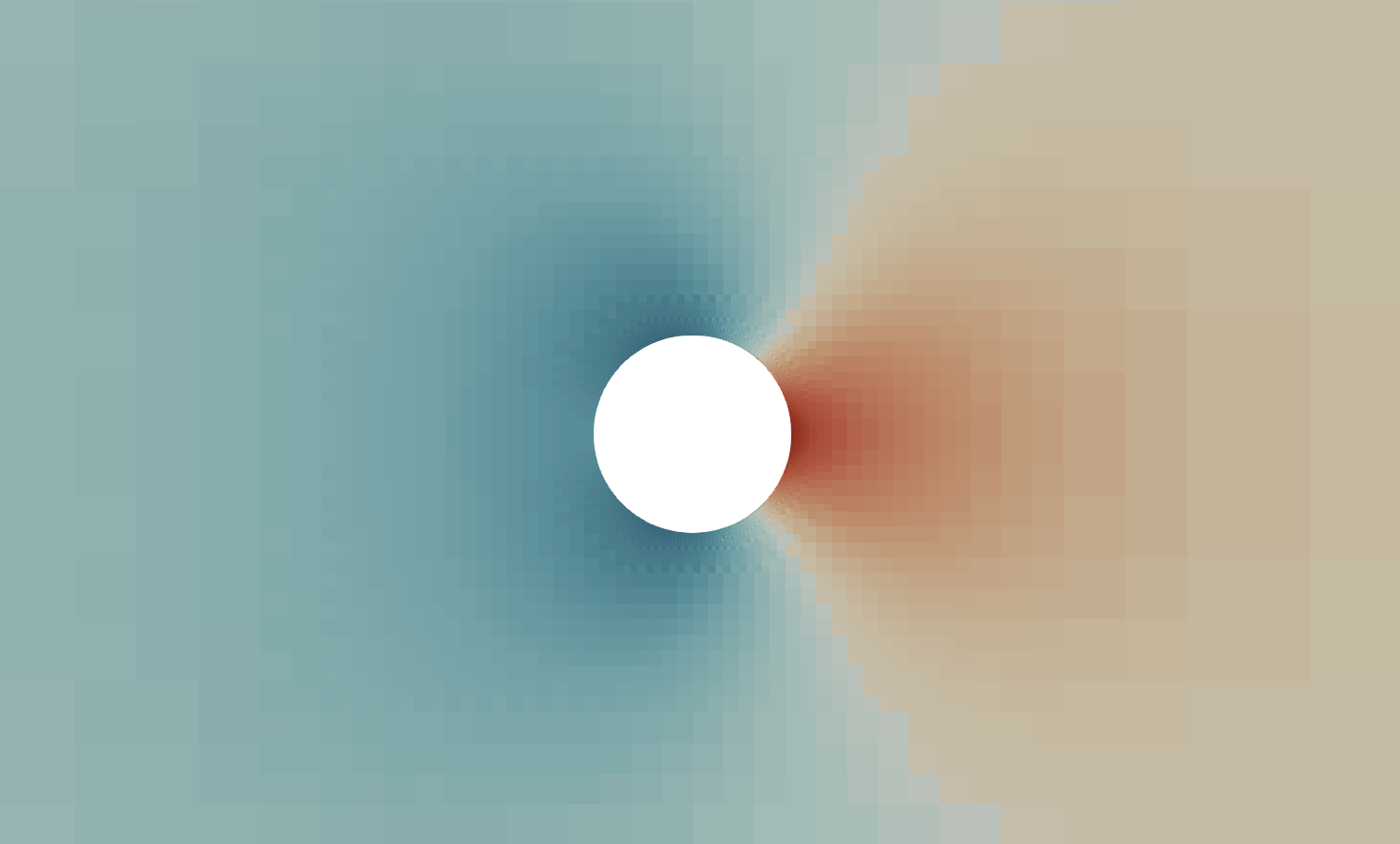}
}
\iftoggle{tikzExternal}{
\input{./tikz/legend_primal_pressure.tikz}
}{
\includegraphics{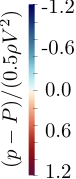}
}
\caption{Submerged cylinder case ($\mathrm{Re}_\mathrm{D} = 10$): Exemplary primal pressure fields. Left to right: classical MWI without relaxation contributions for $\omega = 0.8$ and $\omega = 0.06$, and proposed MWI with relaxation-consistent formulation for $\omega = 0.8$ and $\omega = 0.06$.}
\label{fig:pressure_primal}
\end{figure}

\begin{figure}[!ht]
\centering
\subfigure[]{
\includegraphics[width=0.20\textwidth, trim=10cm 3cm 10cm 3cm, clip]{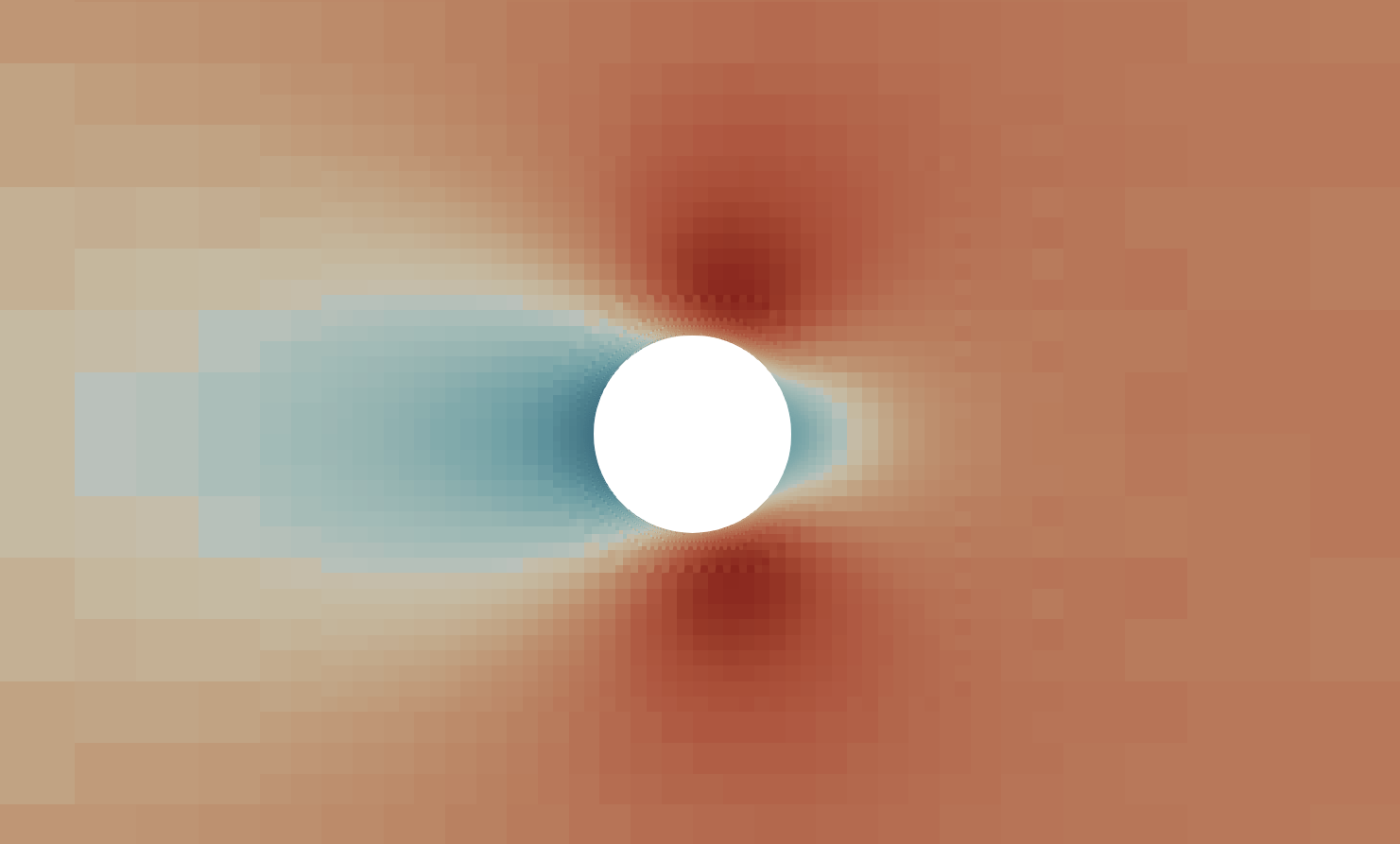}
}
\subfigure[]{
\includegraphics[width=0.20\textwidth, trim=10cm 3cm 10cm 3cm, clip]{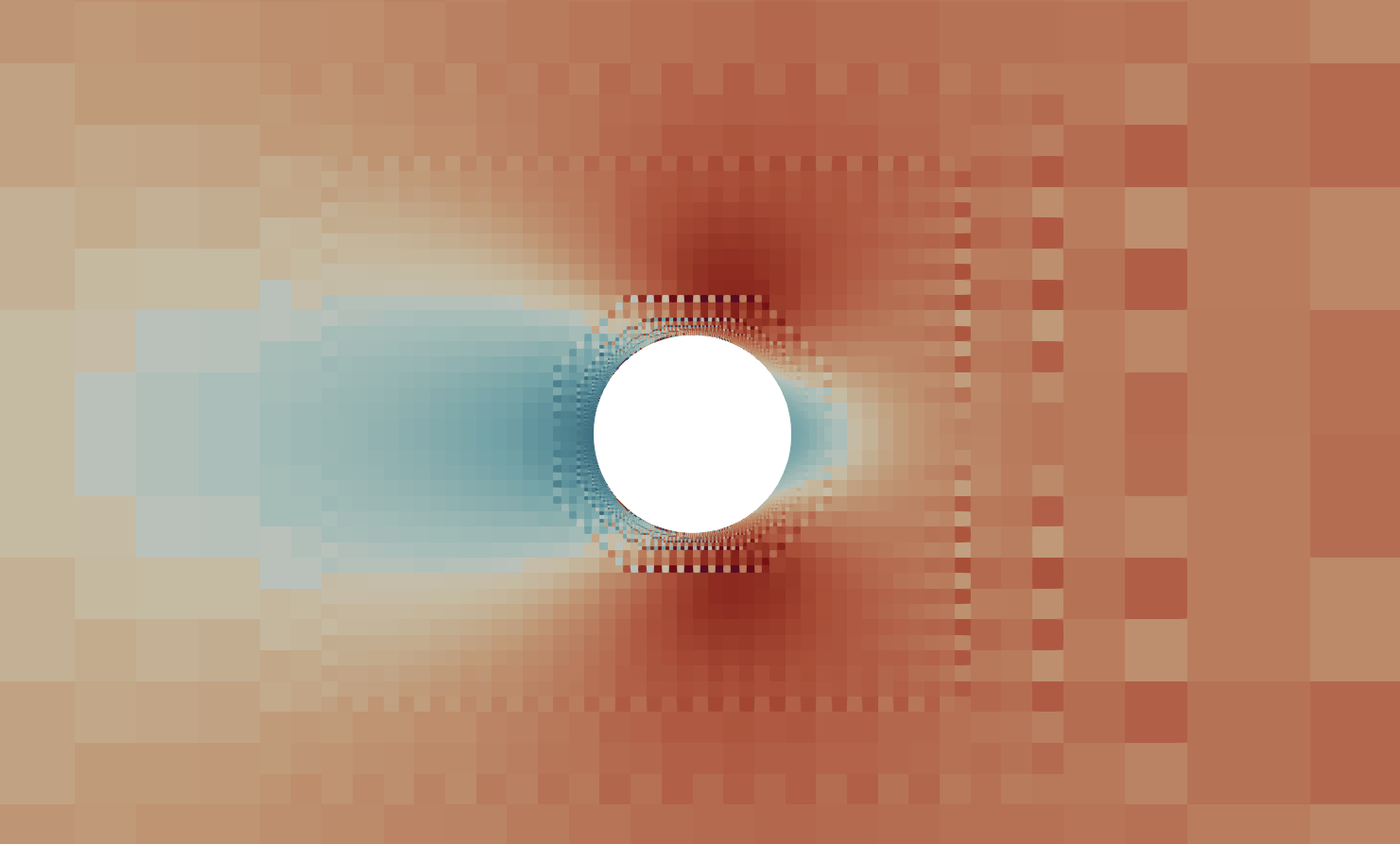}
}
\subfigure[]{
\includegraphics[width=0.20\textwidth, trim=10cm 3cm 10cm 3cm, clip]{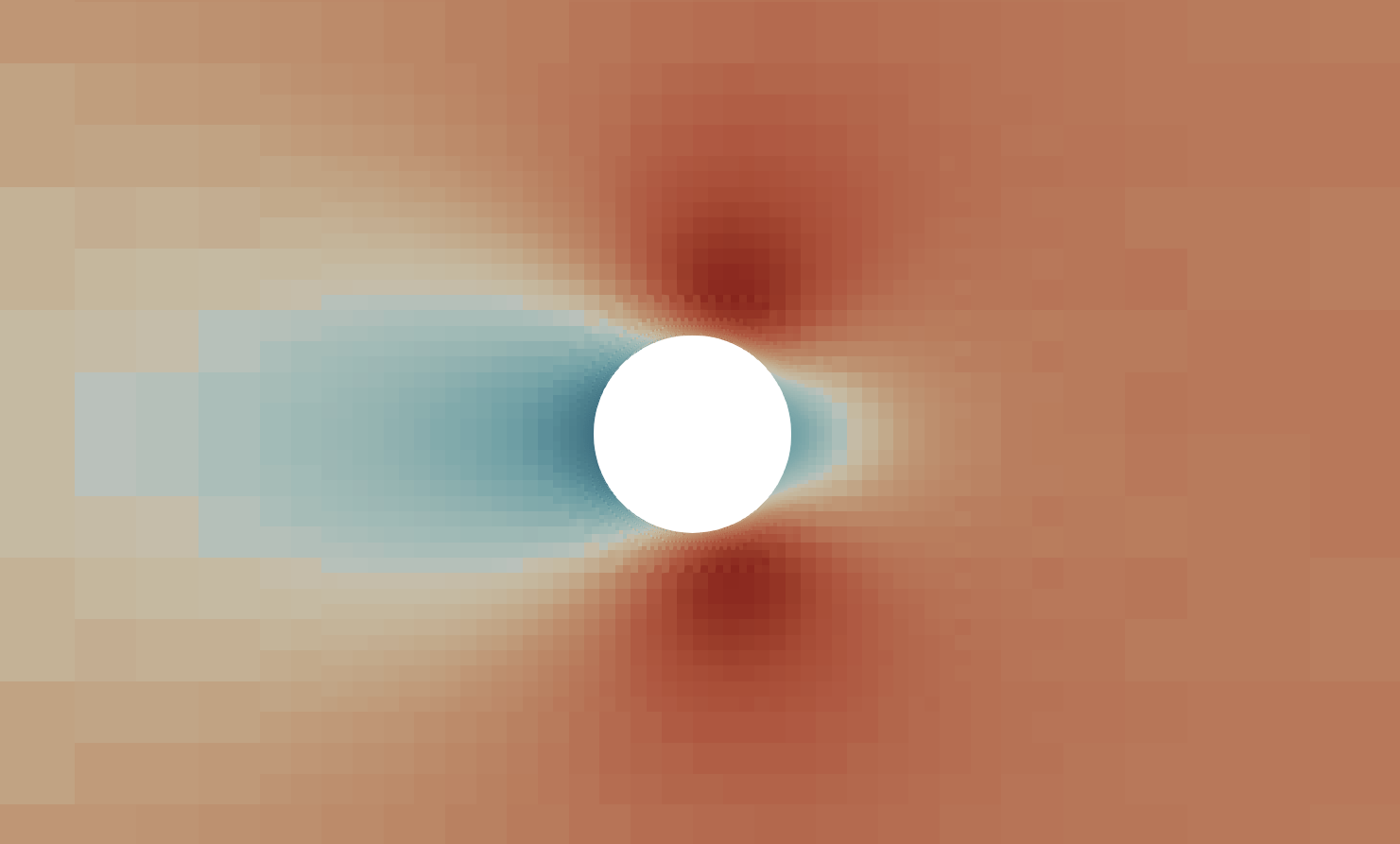}
}
\subfigure[]{
\includegraphics[width=0.20\textwidth, trim=10cm 3cm 10cm 3cm, clip]{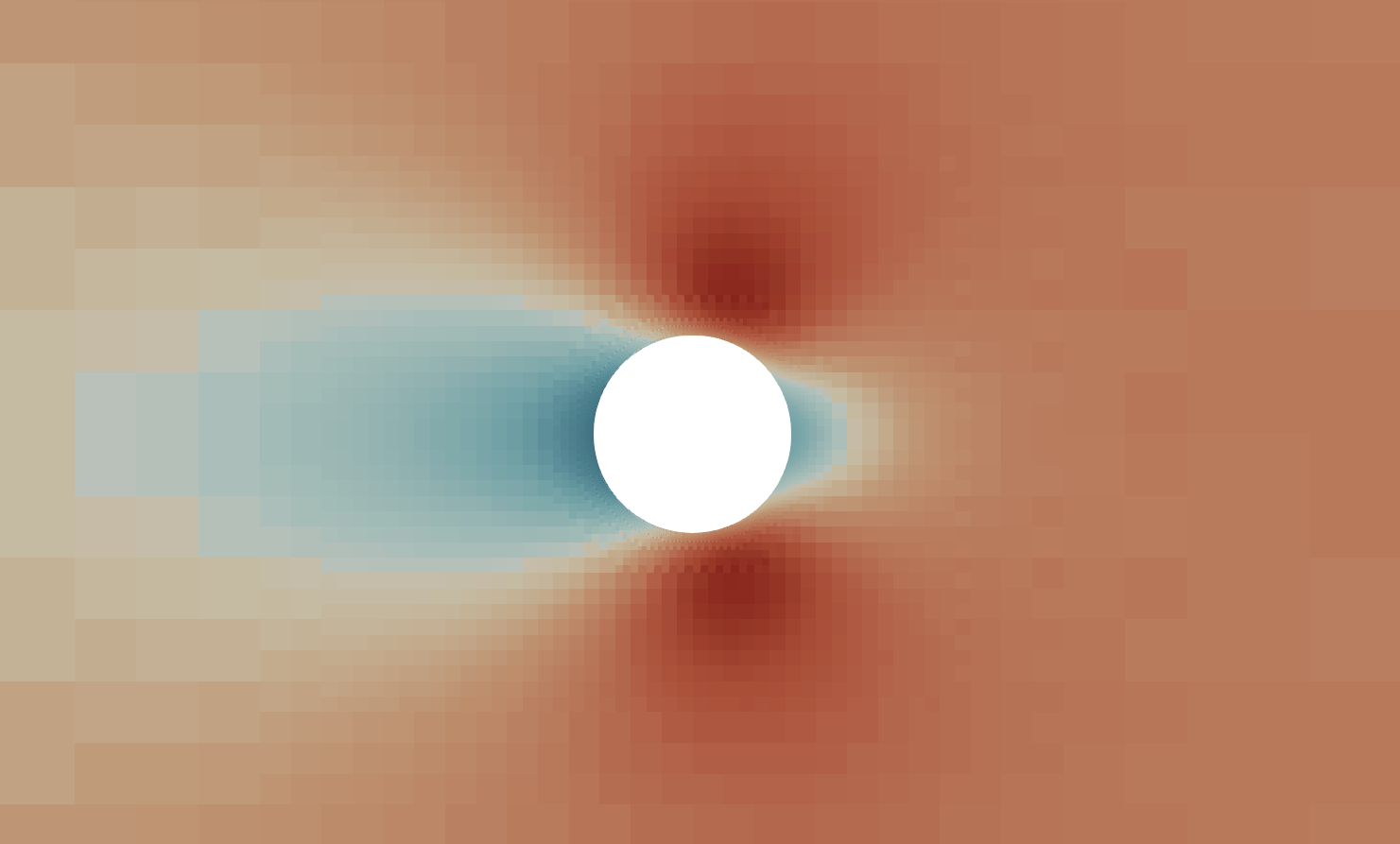}
}
\iftoggle{tikzExternal}{
\input{./tikz/legend_adjoint_pressure.tikz}
}{
\includegraphics{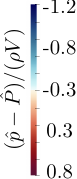}
}
\caption{Submerged cylinder case ($\mathrm{Re}_\mathrm{D} = 10$): Exemplary adjoint pressure fields. Left to right: classical MWI without relaxation contributions for $\omega = 0.8$ and $\omega = 0.06$, and proposed MWI with relaxation-consistent formulation for $\omega = 0.8$ and $\omega = 0.06$.}
\label{fig:pressure_adjoint}
\end{figure}

Exemplary normalized primal pressure fields corresponding to the steady simulations are shown in Fig.~\ref{fig:pressure_primal} for selected relaxation factors and MWI formulations, illustrating the influence of solver-induced contributions on the pressure distribution. Among the four cases, subfigure (b), corresponding to a small relaxation factor without the proposed correction, i.e., Experiment E1.1, exhibits a clearly oscillatory pressure field.
Analogous to the primal pressure fields, the corresponding normalized adjoint pressure distributions are shown in Fig.~\ref{fig:pressure_adjoint}.

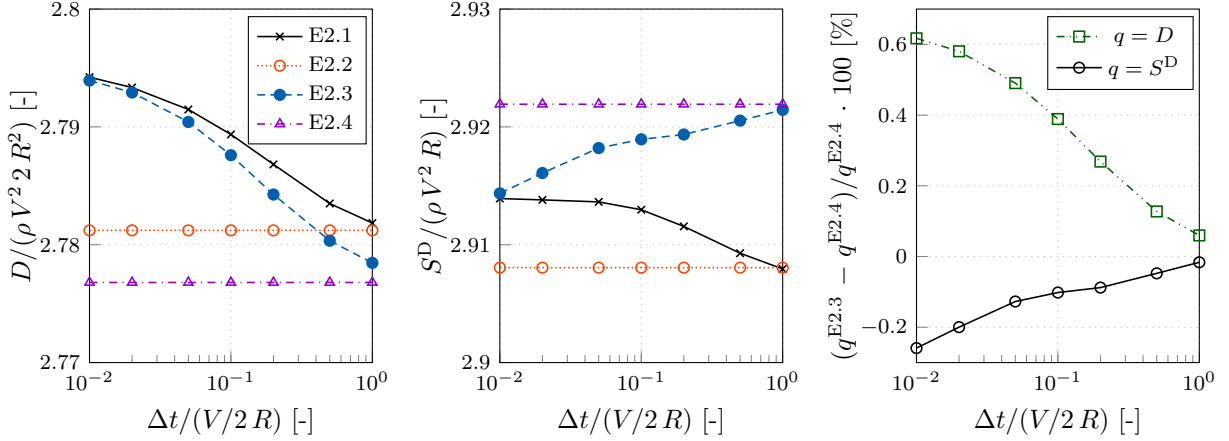
\begin{figure}[!ht]
\centering
\iftoggle{tikzExternal}{
\analytiSolutionPictures
\begin{tikzpicture}
\begin{axis}[
 xlabel={$\Delta t /(V/2\,R)$ [-]},
 xlabel style={text width=0.25\textwidth,align=center},
 ylabel={$D/(\rho \, V^2 \, 2 \, R^2)$ [-]},
 ylabel shift = -2mm,
 ylabel style={text width=0.35\textwidth,align=center},
 legend style={at={(0.98,0.98)},anchor=north east},
 xmode=log,
 xmin=1E-02,
 xmax=1E-00,
 ymin=2.77,
 ymax=2.80,
]

\addplot [line1, mark1, each nth point=1] table[x expr={\thisrowno{0}},y expr={\thisrowno{1}}]{data/cylinder_results_unsteady.dat};
\addplot [line2, mark2, each nth point=1] table[x expr={\thisrowno{0}},y expr={\thisrowno{2}}]{data/cylinder_results_unsteady.dat};
\addplot [line3, mark3, each nth point=1] table[x expr={\thisrowno{0}},y expr={\thisrowno{3}}]{data/cylinder_results_unsteady.dat};
\addplot [line4, mark4, each nth point=1] table[x expr={\thisrowno{0}},y expr={\thisrowno{4}}]{data/cylinder_results_unsteady.dat};

\addlegendentry{E2.1};
\addlegendentry{E2.2};
\addlegendentry{E2.3};
\addlegendentry{E2.4};

\end{axis}
\end{tikzpicture}
\begin{tikzpicture}
\begin{axis}[
 xlabel={$\Delta t /(V/2\,R)$ [-]},
 xlabel style={text width=0.25\textwidth,align=center},
 ylabel={$S^\mathrm{D}/(\rho \, V^2 \, R)$ [-]},
 ylabel shift = -2mm,
 ylabel style={text width=0.35\textwidth,align=center},
 legend style={at={(0.98,0.98)},anchor=north east},
 xmode=log,
 xmin=1E-02,
 xmax=1E-00,
 ymin=2.90,
 ymax=2.93,
]

\addplot [line1, mark1, each nth point=1] table[x expr={\thisrowno{0}},y expr={\thisrowno{5}-6E-03}]{data/cylinder_results_unsteady.dat};
\addplot [line2, mark2, each nth point=1] table[x expr={\thisrowno{0}},y expr={\thisrowno{6}}]{data/cylinder_results_unsteady.dat};
\addplot [line3, mark3, each nth point=1] table[x expr={\thisrowno{0}},y expr={\thisrowno{7}-8E-03}]{data/cylinder_results_unsteady.dat};
\addplot [line4, mark4, each nth point=1] table[x expr={\thisrowno{0}},y expr={\thisrowno{8}}]{data/cylinder_results_unsteady.dat};


\end{axis}
\end{tikzpicture}
\begin{tikzpicture}
\begin{axis}[
 xlabel={$\Delta t /(V/2\,R)$ [-]},
 xlabel style={text width=0.25\textwidth,align=center},
 ylabel={$(q^\mathrm{E2.3} - q^\mathrm{E2.4})/q^\mathrm{E2.4} \cdot 100$ [\%]},
 ylabel shift = -2mm,
 ylabel style={text width=0.35\textwidth,align=center},
 legend style={at={(0.98,0.98)},anchor=north east},
 xmode=log,
 xmin=1E-02,
 xmax=1E-00,
 ymin=-0.3,
 ymax=0.7,
]

\addplot [line5, mark5, each nth point=1] table[x expr={\thisrowno{0}},y expr={ ((\thisrowno{3}-\thisrowno{4})/\thisrowno{4})*100 }]{data/cylinder_results_unsteady.dat};
\addplot [line6, mark6, each nth point=1] table[x expr={\thisrowno{0}},y expr={ ((\thisrowno{7}-8E-03-\thisrowno{8})/\thisrowno{8})*100 }]{data/cylinder_results_unsteady.dat};

\addlegendentry{$q = D$};
\addlegendentry{$q = S^\mathrm{D}$};

\end{axis}
\end{tikzpicture}
}{
\includegraphics{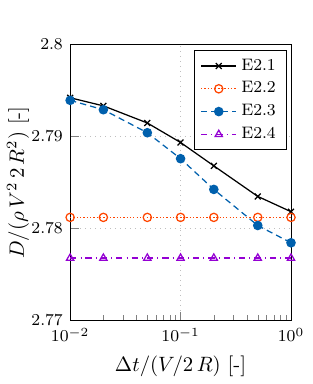}
\includegraphics{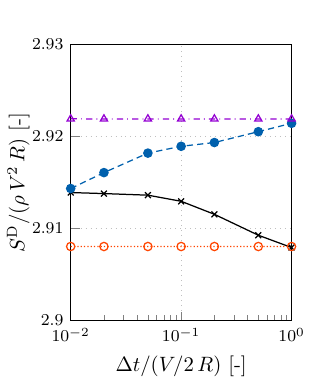}
\includegraphics{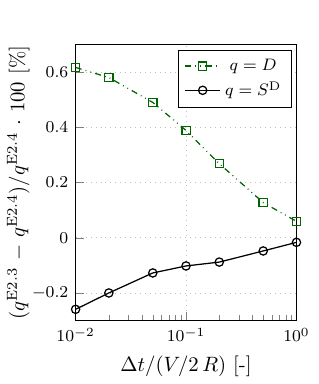}
}
\caption{Submerged cylinder case ($\mathrm{Re}_\mathrm{D} = 10$): Influence of the normalized pseudo-time step $\Delta t / (V/D)$ on the primal drag coefficient (left) and the corresponding adjoint sensitivity (center) for Experiments E2.1--E2.4. The right-hand side shows the relative deviation between the formulations.}
\label{fig:cylinder_unsteady_results}
\end{figure}

The corresponding results for the unsteady simulations are shown in Fig.~\ref{fig:cylinder_unsteady_results}, where the primal drag coefficient (left) and the corresponding adjoint sensitivity (center) are plotted over the normalized pseudo-time step size $\Delta t / (V/D)$ for the configurations E2.1--E2.4. As expected, two of the curves remain constant, while the other two exhibit a clear dependence on the pseudo-time step size. The constant predictions correspond to the formulations that include temporal contributions, i.e., E2.2 and E2.4, which differ only in their treatment of relaxation, resulting in a constant offset between the curves. The relaxation parameter also introduces additional differences between the non-constant predictions of E2.1 and E2.3. These formulations show a strong deviation for small time-step sizes and a weaker deviation for larger time steps, consistent with the observations from the steady relaxation study. The predictions approach the constant values of the time-consistent formulations as the time-step size increases, but still deviate slightly at $\Delta t/(V/D) = 1$. The behavior of the predicted drag coefficient and sensitivity is generally similar; however, in case E2.3, the solution approaches the constant value from below rather than from above. This explains the negative deviation observed in the right-hand plot, where the relative difference $(q^{\mathrm{E2.3}} - q^{\mathrm{E2.4}})/q^{\mathrm{E2.4}} \cdot 100$ [\%] between E2.3 and E2.4 is shown for both quantities. In this case, the influence on the primal drag coefficient reaches approximately $+0.6\%$, which is larger than the deviation in the adjoint sensitivity of about $-0.2\%$.

A repeated visualization of the corresponding pressure fields is omitted, as the observed patterns are qualitatively identical to those of the steady case.


\subsection{Three-Dimensional Ship Hull Flow}
To further assess the influence of solver-induced contributions on the MWI in a more complex setting, a three-dimensional turbulent flow around a ship hull is considered. The test case is based on the Japan Bulk Carrier, which was extensively investigated within the Tokyo 2015 CFD Workshop, cf. \cite{hino2020numerical}. The Reynolds number is defined as $\mathrm{Re}_\mathrm{L} = \rho V L / \mu = 1 \times 10^{7}$, where $L$ denotes the ship length overall submerged (LOS). The velocity $V$ corresponds to the model-scale towing speed, while $\rho$ and $\mu$ denote the density and dynamic viscosity of water, respectively. The simulations are performed at model scale with a scale factor of $\lambda = 32.48$, corresponding to a full-scale cruising speed of $14.5\,\mathrm{kn}$.

All simulations are performed in double-body mode, i.e., without a free surface and with a symmetry condition imposed at the undisturbed water plane. The computational domain extends from $x_\mathrm{1}/L = -2$ to $x_\mathrm{1}/L = 2$ in streamwise direction, from $x_\mathrm{2}/L = 0$ to $x_\mathrm{2}/L = 1$ in the lateral direction, and from $x_\mathrm{3}/L = -1$ to $x_\mathrm{3}/L = 0.05$ in the vertical direction, where the latter defines the vessel's draft. The coordinate origin is located at the aft perpendicular at keel level. For resistance computations, only half of the hull is considered, whereas for propulsion simulations, the full geometry is used. A uniform velocity is prescribed at the inlet, while a pressure outlet condition is applied downstream. Symmetry conditions are imposed at the undisturbed water plane and, in the case of resistance simulations, additionally at the ship centerline plane. Velocity boundary conditions are applied at the bottom as well as on the remaining far-field boundaries.

\begin{figure}[!ht]
\centering
\subfigure[]{
\includegraphics[width=0.9\textwidth]{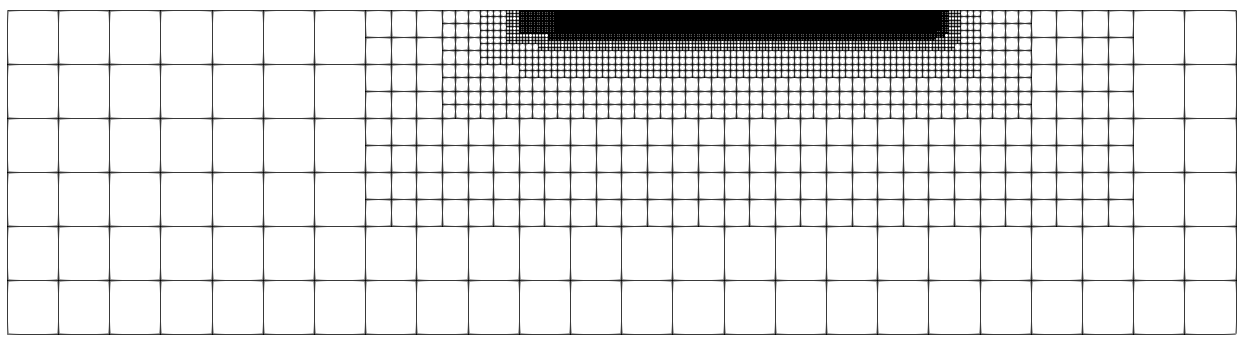}
}
\subfigure[]{
\includegraphics[width=0.9\textwidth]{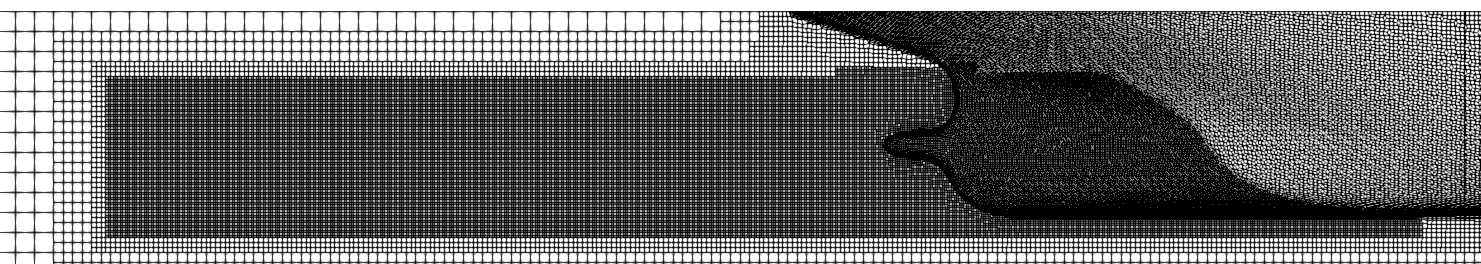}
}
\caption{Japan Bulk Carrier case ($\mathrm{Re}_\mathrm{L} = 10^{7}$): Unstructured computational grid. Far-field (a) and near-hull view (b) on the utilized numerical grid.}
\label{fig:ship_grid}
\end{figure}

An impression of the unstructured computational grid for half the model, consisting of approximately \SI{2 900 000}{} control volumes, is provided in Fig.~\ref{fig:ship_grid}.

Two groups of numerical experiments are conducted:
\begin{itemize}
    \item \textbf{E1.1:} Steady simulations are carried out on a half-domain configuration exploiting symmetry with respect to the center plane. The relaxation factor $\omega$ is varied over the range $6\times10^{-2}$ to $1\times10^{-8}$. Results are obtained using the classical MWI formulation from Eqn.~\ref{equ:mwi_final}.
    
    \item \textbf{E1.2:} The same steady simulations are performed using the proposed MWI formulation from Eqn.~\ref{equ:mwi_relax_final}, allowing for a direct comparison of solver-induced effects.
    
    \item \textbf{E2.1:} Pseudo-unsteady simulations are performed on the full ship geometry in self-propulsion mode. Propulsion is modeled using an actuator disk approach, in which the propulsive force is introduced as a volumetric source term based on a prescribed open-water curve. The normalized time-step size $\Delta t/(L/V)$ is varied over the range $10^{-2}$ to $10^{1}$ using the classical formulation  from Eqn.~\ref{equ:mwi_final}.
    
    \item \textbf{E2.2:} The same unsteady simulations are carried out using the proposed MWI formulation from Eqn.~\ref{equ:mwi_time_final}, enabling a consistent comparison with respect to temporal effects.
\end{itemize}

The considered objective functional are defined as follows: $D$ denotes the total resistance (drag), $W$ the wake homogeneity in the absence of propulsion, i.e., the inhomogeneity of the nominal wake, $T$ the propeller thrust, and $E$ the wake homogeneity in the presence of an actuator disk, i.e., the inhomogeneity of the propeller-influenced (effective) wake field.
In line with Eqn.~\ref{equ:cylinder_qoi}, which is extended to effective quantities due to the employed RANS approach, i.e., $p \to p^\mathrm{eff}$ and $\mu \to \mu^\mathrm{eff}$, the wake homogeneity functionals $W$ and $E$ are defined analogously as
\begin{align}
    W,\,E = \frac{1}{\Delta \Omega^\mathrm{Disc} \, \bar{v}_\mathrm{axial}} \int \frac{1}{2} \left( v_\mathrm{axial} - \bar{v}_\mathrm{axial} \right)^2 \, \mathrm{d} \Omega^\mathrm{Disc} \, ,
\end{align}
where $\Delta \Omega^\mathrm{Disc}$ denotes a thin disk-shaped control volume in which the wake is evaluated and, in the case of propulsion simulations, body forces are applied. The quantity $v_\mathrm{axial}$ denotes the local velocity component in the streamwise (axial) direction, and $\bar{v}_\mathrm{axial}$ represents the corresponding spatial average within this region.
The corresponding shape sensitivity expressions remain identical to Eqn.~\ref{equ:cylinder_qoi} and are denoted by $S^\mathrm{D}$, $S^\mathrm{T}$, $S^\mathrm{W}$, and $S^\mathrm{E}$ for drag, thrust, and wake homogeneity in nominal and effective conditions, respectively. Further possibilities to consider wake objectives from an adjoint perspective are discussed in  \cite{stuck2011adjointII}.

All adjoint computations are performed based on a fixed primal solution obtained with the solver-independent formulation, thereby excluding any influence of relaxation-induced effects on the reference flow field.
In the case of the adjoint wake-homogeneity functional, additional adjoint momentum source terms are present. Accordingly, the MWI formulation including body-force contributions, as introduced in \cite{kuhl2022discrete}, is employed throughout.

\begin{figure}[!ht]
\centering
\iftoggle{tikzExternal}{
\analytiSolutionPictures
\begin{tikzpicture}
\begin{axis}[
 xlabel={$\omega$ [-]},
 xlabel style={text width=0.25\textwidth,align=center},
 ylabel={$2 \, D/(\rho \, V^2 \, L^2) \cdot 10^{4}$ [-]},
 ylabel shift = -2mm,
 ylabel style={text width=0.25\textwidth,align=center},
 legend style={at={(0.98,0.02)},anchor=south east},
 xmode=log,
 xmin=6E-02,
 xmax=1E-00,
 ymin=4.26,
 ymax=4.36,
]

\addplot [line1, mark1, each nth point=1] table[x expr={\thisrowno{0}},y expr={\thisrowno{1}*1E+04}]{data/jbc_results_steady.dat};
\addplot [line2, mark2, each nth point=1] table[x expr={\thisrowno{0}},y expr={\thisrowno{2}*1E+04}]{data/jbc_results_steady.dat};

\addlegendentry{E1.1};
\addlegendentry{E1.2};

\end{axis}
\end{tikzpicture}
\begin{tikzpicture}
\begin{axis}[
 xlabel={$\omega$ [-]},
 xlabel style={text width=0.25\textwidth,align=center},
 ylabel={$2 \, S^\mathrm{D}/(\rho \, V^2 \, L) \cdot 10^{2}$ [-]},
 ylabel shift = -2mm,
 ylabel style={text width=0.25\textwidth,align=center},
 legend style={at={(0.98,0.02)},anchor=south east},
 xmode=log,
 xmin=6E-02,
 xmax=1E-00,
 ymin=-4.80,
 ymax=-4.72,
]

\addplot [line3, mark3, each nth point=1] table[x expr={\thisrowno{0}},y expr={\thisrowno{5}*1E+02}]{data/jbc_results_steady.dat};
\addplot [line4, mark4, each nth point=1] table[x expr={\thisrowno{0}},y expr={\thisrowno{6}*1E+02}]{data/jbc_results_steady.dat};

\addlegendentry{E1.1};
\addlegendentry{E1.2};

\end{axis}
\end{tikzpicture}
\begin{tikzpicture}
\begin{axis}[
 xlabel={$\omega$ [-]},
 xlabel style={text width=0.25\textwidth,align=center},
 ylabel={$(q^\mathrm{E1.1} - q^\mathrm{E1.2})/q^\mathrm{E1.2} \cdot 100$ [\%]},
 ylabel shift = -2mm,
 ylabel style={text width=0.25\textwidth,align=center},
 legend style={at={(0.98,0.98)},anchor=north east},
 xmode=log,
 xmin=6E-02,
 xmax=1E-00,
 ymin=-1.5,
 ymax=1.5,
]

\addplot [line5, mark5, each nth point=1] table[x expr={\thisrowno{0}},y expr={ ((\thisrowno{1}-\thisrowno{2})/\thisrowno{2})*100 }]{data/jbc_results_steady.dat};
\addplot [line6, mark6, each nth point=1] table[x expr={\thisrowno{0}},y expr={ ((\thisrowno{5}-\thisrowno{6})/\thisrowno{6})*100 }]{data/jbc_results_steady.dat};

\addlegendentry{$q = D$};
\addlegendentry{$q = S^\mathrm{D}$};

\end{axis}
\end{tikzpicture}
}{
\includegraphics{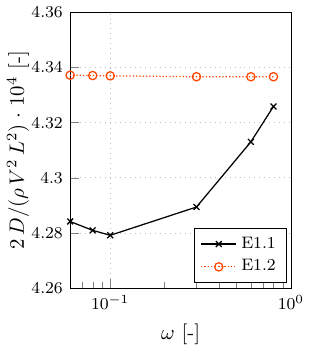}
\includegraphics{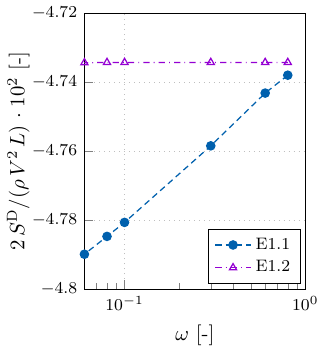}
\includegraphics{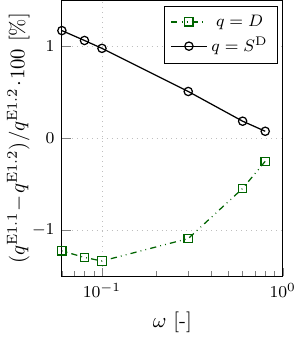}
}
\caption{Japan Bulk Carrier case ($\mathrm{Re} = 10^{7}$): Influence of the relaxation factor $\omega$ on the normalized total resistance (left) and the corresponding normalized drag sensitivity (center) for Experiments E1.1 and E1.2. The right-hand side shows the relative deviation between both formulations.}
\label{fig:jbc_steady_results_drag}
\end{figure}
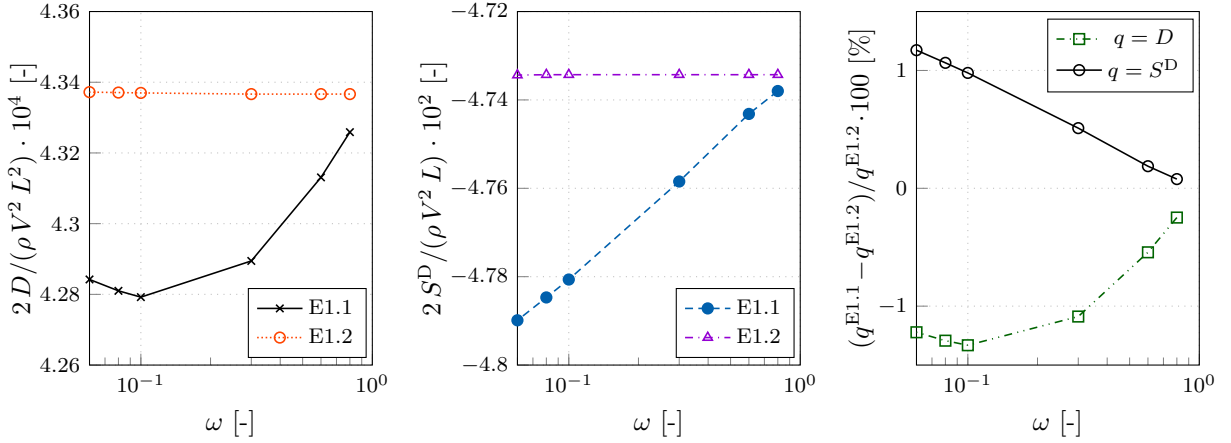

The corresponding results for the steady simulations are shown in Fig.~\ref{fig:jbc_steady_results_drag}. The normalized total resistance (left) and the corresponding normalized drag sensitivity (center) are plotted over the relaxation factor $\omega$ for the classical (E1.1) and the proposed MWI formulation (E1.2). It is observed that both quantities remain constant for the proposed formulation, serving as a reference solution, whereas the classical formulation exhibits a clear dependence on the relaxation parameter. The deviation from the reference almost always increases for decreasing values of $\omega$ and vanishes only in the limit $\omega \to 1$. The relative differences between the two formulations are summarized in the right-hand plot. The maximum deviations amount to approximately $-1\%$ for the resistance and $+1\%$ for the corresponding shape sensitivity.

\begin{figure}[!ht]
\centering
\iftoggle{tikzExternal}{
\analytiSolutionPictures
\begin{tikzpicture}
\begin{axis}[
 xlabel={$\omega$ [-]},
 xlabel style={text width=0.25\textwidth,align=center},
 ylabel={$W \cdot 10$ [-]},
 ylabel shift = -2mm,
 ylabel style={text width=0.35\textwidth,align=center},
 legend style={at={(0.98,0.98)},anchor=north east},
 xmode=log,
 xmin=6E-02,
 xmax=1E-00,
 ymin=0.1,
 ymax=0.14,
 scaled y ticks = false,
]

\addplot [line1, mark1, each nth point=1] table[x expr={\thisrowno{0}},y expr={\thisrowno{3}}]{data/jbc_results_steady.dat};
\addplot [line2, mark2, each nth point=1] table[x expr={\thisrowno{0}},y expr={\thisrowno{4}}]{data/jbc_results_steady.dat};

\addlegendentry{E1.1};
\addlegendentry{E1.2};

\end{axis}
\end{tikzpicture}
\begin{tikzpicture}
\begin{axis}[
 xlabel={$\omega$ [-]},
 xlabel style={text width=0.25\textwidth,align=center},
 ylabel={$S^\mathrm{W} L \cdot 10^{-2}$ [-]},
 ylabel shift = -2mm,
 ylabel style={text width=0.35\textwidth,align=center},
 legend style={at={(0.98,0.02)},anchor=south east},
 xmode=log,
 xmin=6E-02,
 xmax=1E-00,
 ymin=-0.8,
 ymax=-0.2,
]

\addplot [line3, mark3, each nth point=1] table[x expr={\thisrowno{0}},y expr={\thisrowno{7}*1E-02}]{data/jbc_results_steady__coarse.dat};
\addplot [line4, mark4, each nth point=1] table[x expr={\thisrowno{0}},y expr={\thisrowno{8}*1E-02}]{data/jbc_results_steady__coarse.dat};

\addlegendentry{E1.1};
\addlegendentry{E1.2};

\end{axis}
\end{tikzpicture}
\begin{tikzpicture}
\begin{axis}[
 xlabel={$\omega$ [-]},
 xlabel style={text width=0.25\textwidth,align=center},
 ylabel={$(q^\mathrm{E1.1} - q^\mathrm{E1.2})/q^\mathrm{E1.2} \cdot 100$ [\%]},
 ylabel shift = -2mm,
 ylabel style={text width=0.35\textwidth,align=center},
 legend style={at={(0.98,0.02)},anchor=south east},
 xmode=log,
 xmin=6E-02,
 xmax=1E-00,
 ymin=-60,
 ymax=40,
]

\addplot [line5, mark5, each nth point=1] table[x expr={\thisrowno{0}},y expr={ ((\thisrowno{3}-\thisrowno{4})/\thisrowno{4})*100 }]{data/jbc_results_steady.dat};
\addplot [line6, mark6, each nth point=1] table[x expr={\thisrowno{0}},y expr={ ((\thisrowno{7}-\thisrowno{8})/\thisrowno{8})*100 }]{data/jbc_results_steady__coarse.dat};

\addlegendentry{$q = W$};
\addlegendentry{$q = S^\mathrm{W}$};

\end{axis}
\end{tikzpicture}
}{
\includegraphics{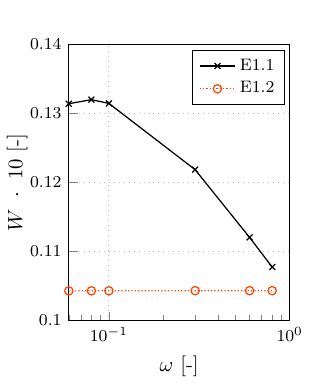}
\includegraphics{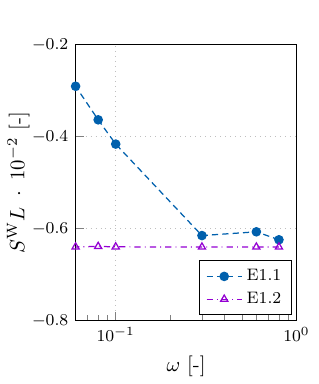}
\includegraphics{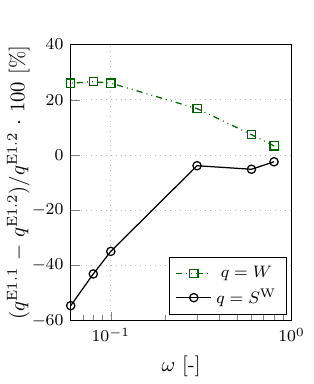}
}
\caption{Japan Bulk Carrier case ($\mathrm{Re} = 10^{7}$): Influence of the relaxation factor $\omega$ on the wake homogeneity $W$ (left) and the corresponding shape sensitivity $S^\mathrm{\omega}$ (center) for Experiments E1.1 and E1.2. The right-hand side shows the relative deviation between both formulations.}
\label{fig:jbc_steady_results_wake}
\end{figure}
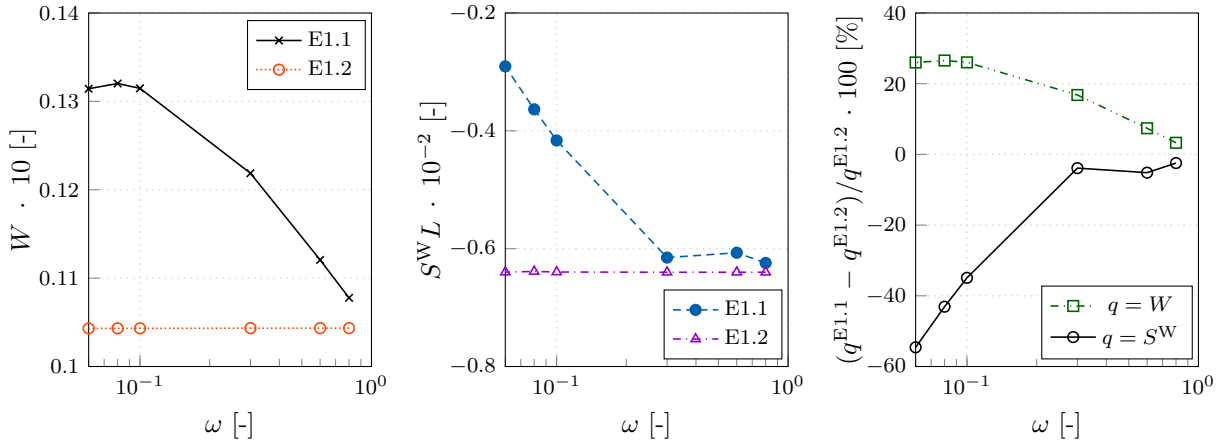

The corresponding nominal wake results for the steady simulations are shown in Fig.~\ref{fig:jbc_steady_results_wake}. Qualitatively, the behavior is similar to the drag-related quantities: the proposed formulation (E1.2) yields results that are independent of the relaxation factor $\omega$, whereas the classical formulation (E1.1) exhibits a pronounced dependency, with increasing deviations for smaller values of $\omega$. 
However, the quantitative differences are significantly larger. For the wake homogeneity $W$ (left), deviations of up to approximately $+25\%$ are observed. The effect is even more pronounced for the corresponding shape sensitivity (center), where a sign change occurs around $\omega \approx 0.5$, leading to negative values and maximum deviations beyond $-100\%$. The relative differences between the two formulations are summarized in the right-hand plot. In particular, the observed sign change can have severe implications for optimization procedures, as it may compromise the computation of consistent descent directions.

\begin{figure}[!ht]
\centering
\subfigure[]{
\includegraphics[height=0.2\textheight, trim=10cm 0cm 27cm 0cm, clip]{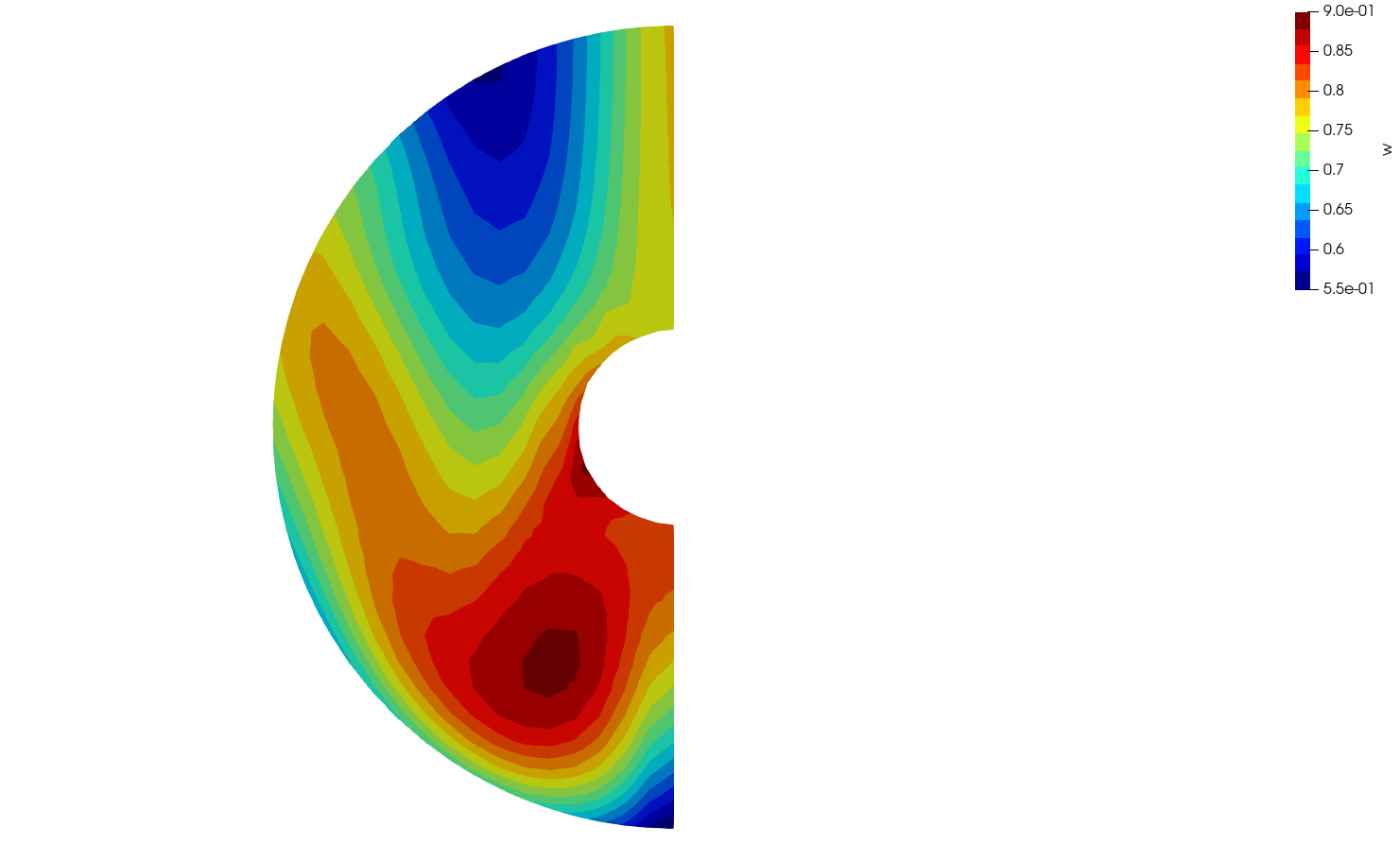}
}
\quad
\subfigure[]{
\includegraphics[height=0.2\textheight, trim=10cm 0cm 27cm 0cm, clip]{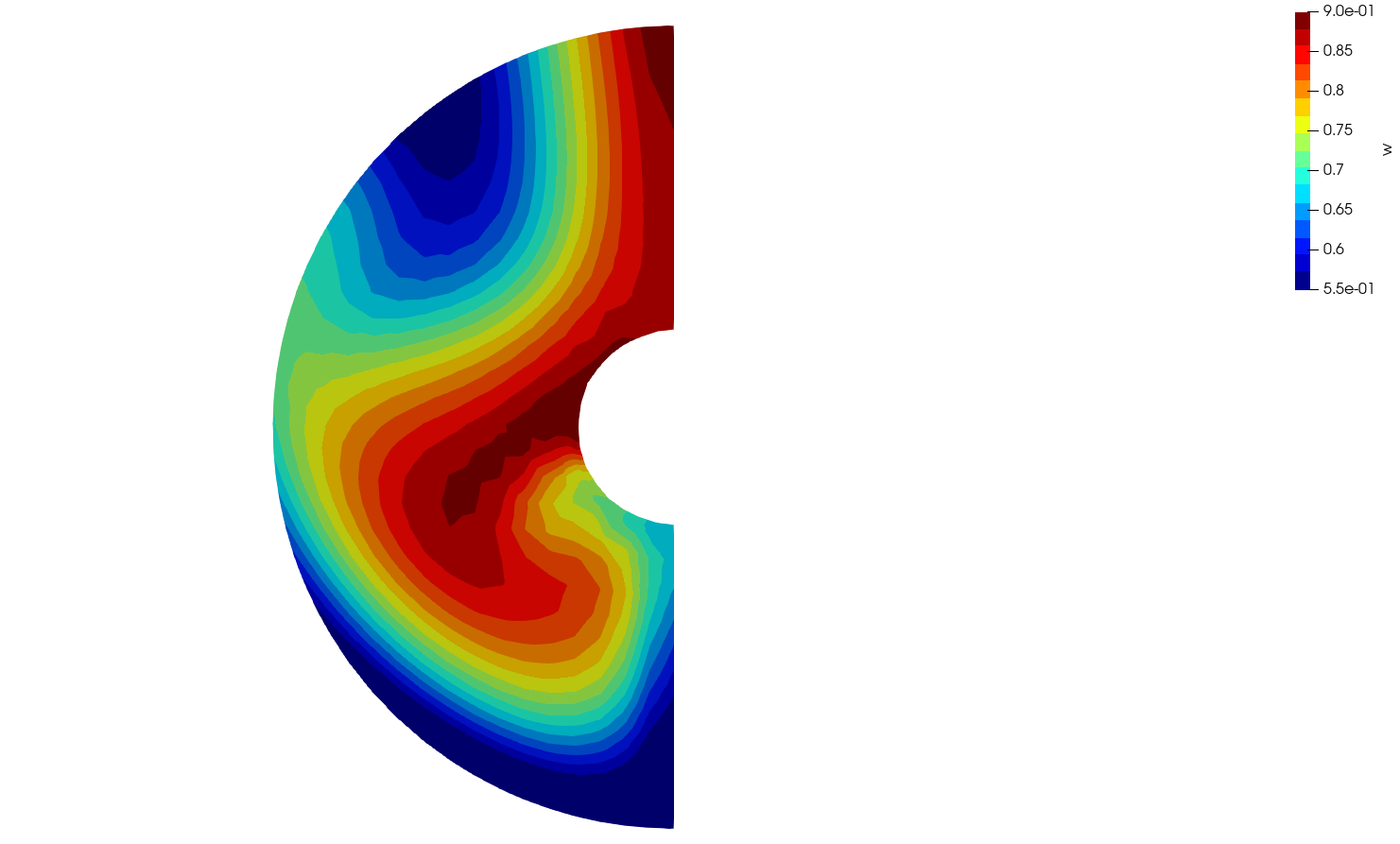}
}
\qquad \qquad
\subfigure[]{
\includegraphics[height=0.2\textheight, trim=10cm 0cm 27cm 0cm, clip]{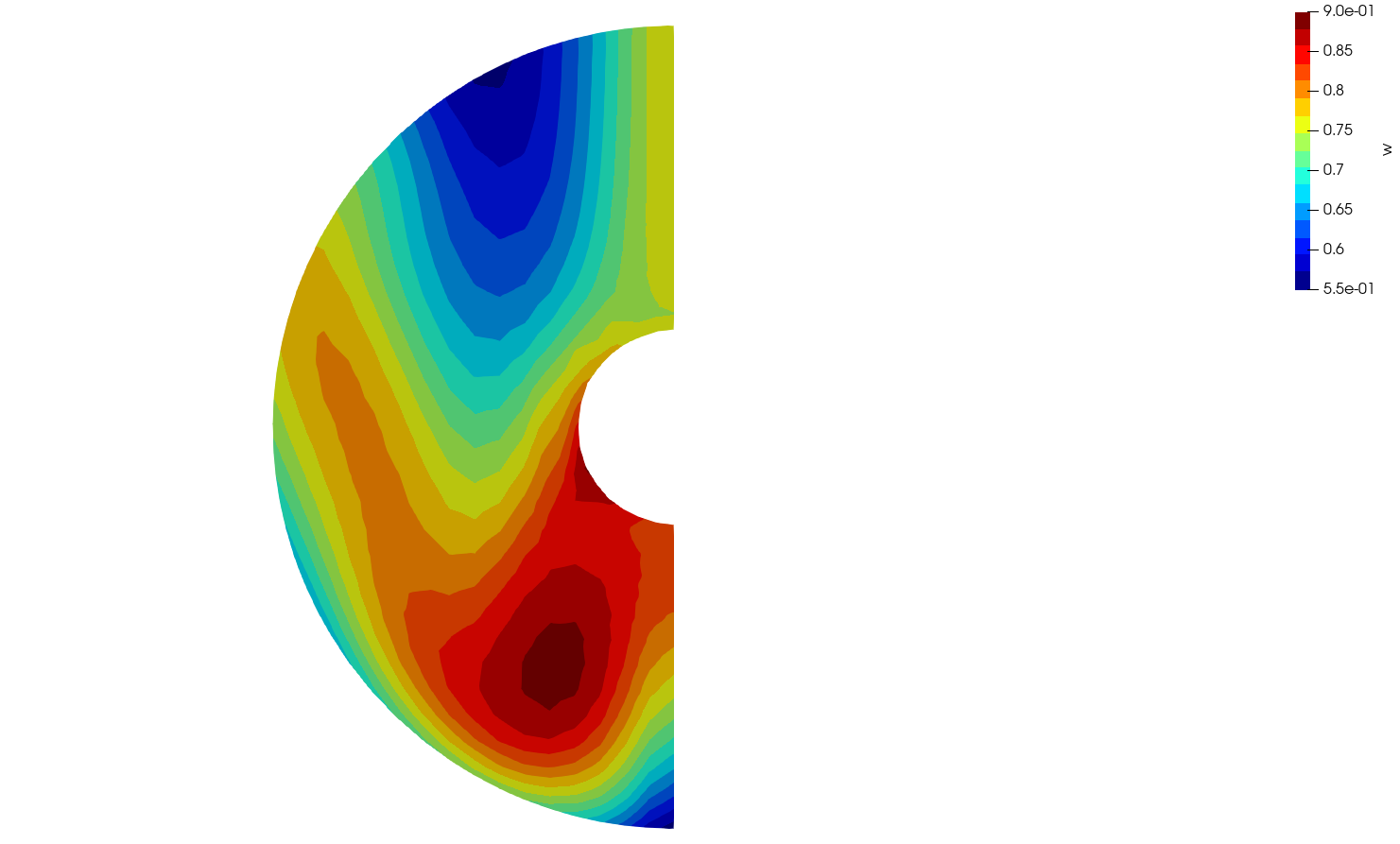}
}
\quad
\subfigure[]{
\includegraphics[height=0.2\textheight, trim=10cm 0cm 27cm 0cm, clip]{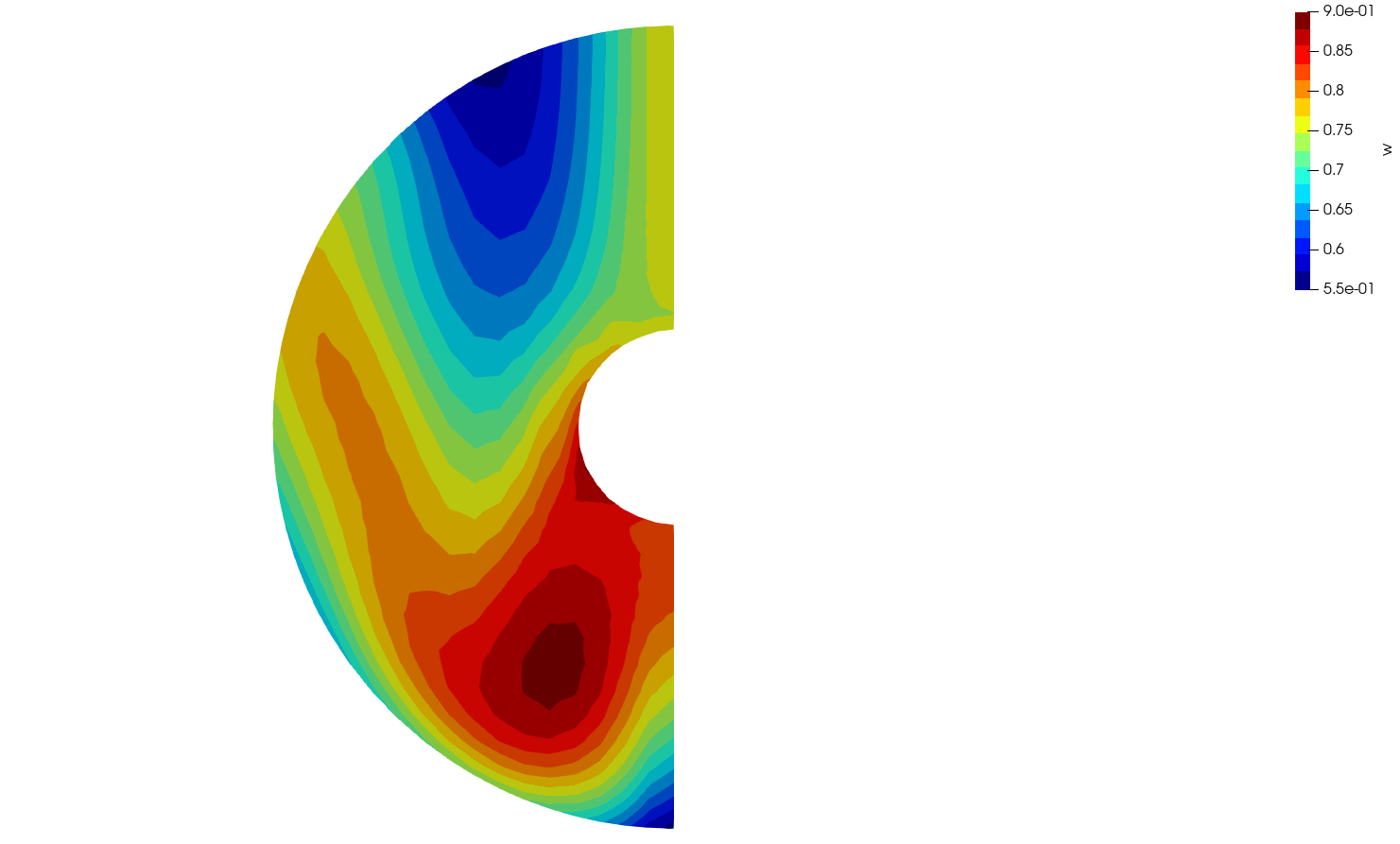}
}
\qquad \qquad
\iftoggle{tikzExternal}{
\input{./tikz/legend_wake.tikz}
}{
\includegraphics{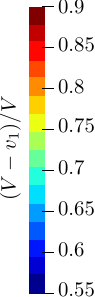}
}
\caption{Japan Bulk Carrier case ($\mathrm{Re} = 10^{7}$): Exemplary nominal wake fields. (a)--(b) Classical MWI formulation without correction for $\omega = 0.8$ and $\omega = 0.06$, respectively; (c)--(d) proposed MWI formulation for the same relaxation factors.}
\label{fig:jbc_wake}
\end{figure}

\begin{figure}[!ht]
\centering
\subfigure[]{
\includegraphics[height=0.2\textheight, trim=18cm 0cm 21cm 0cm, clip]{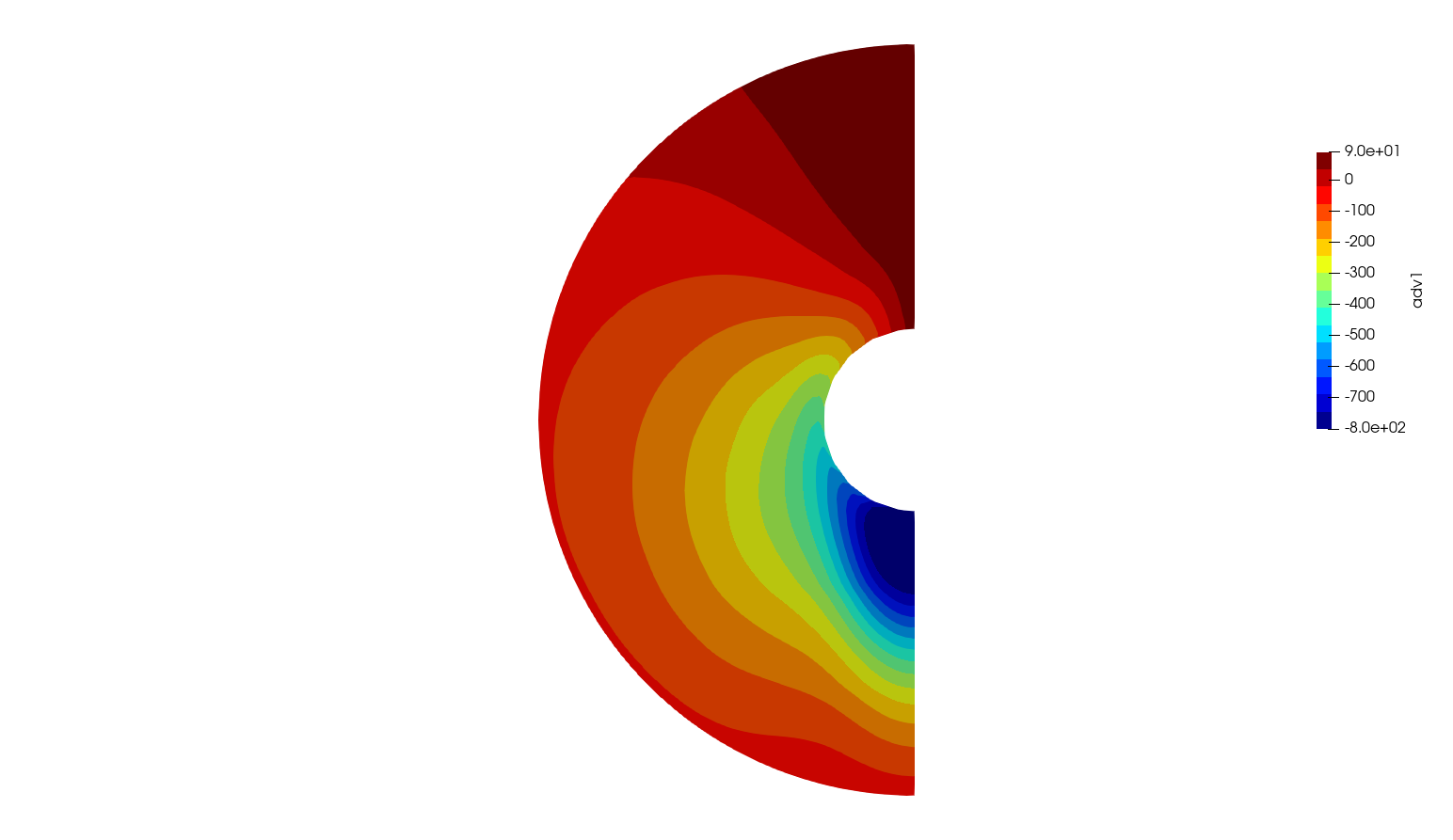}
}
\quad
\subfigure[]{
\includegraphics[height=0.2\textheight, trim=10cm 0cm 28cm 0cm, clip]{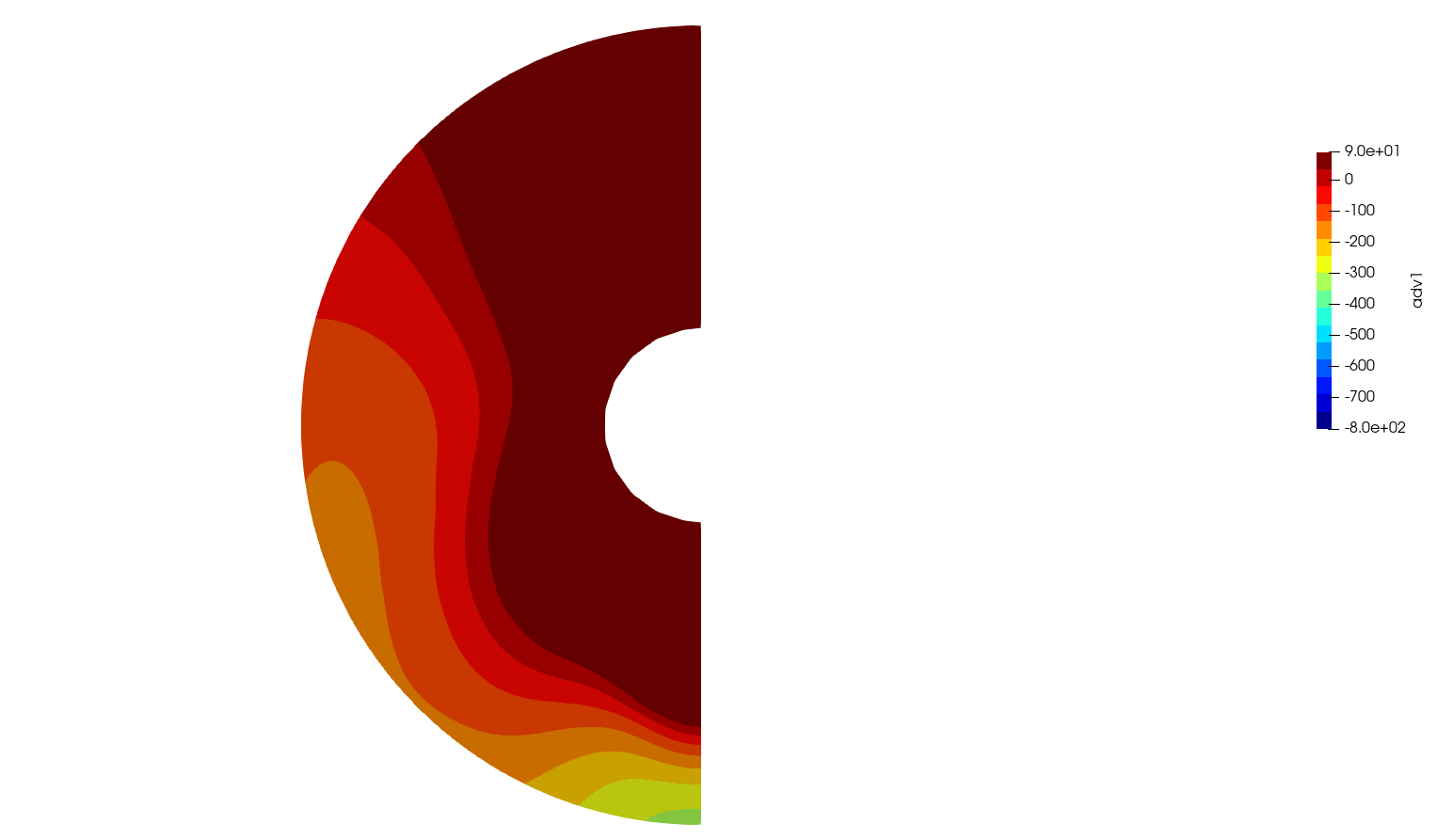}
}
\qquad \qquad
\subfigure[]{
\includegraphics[height=0.2\textheight, trim=10cm 0cm 28cm 0cm, clip]{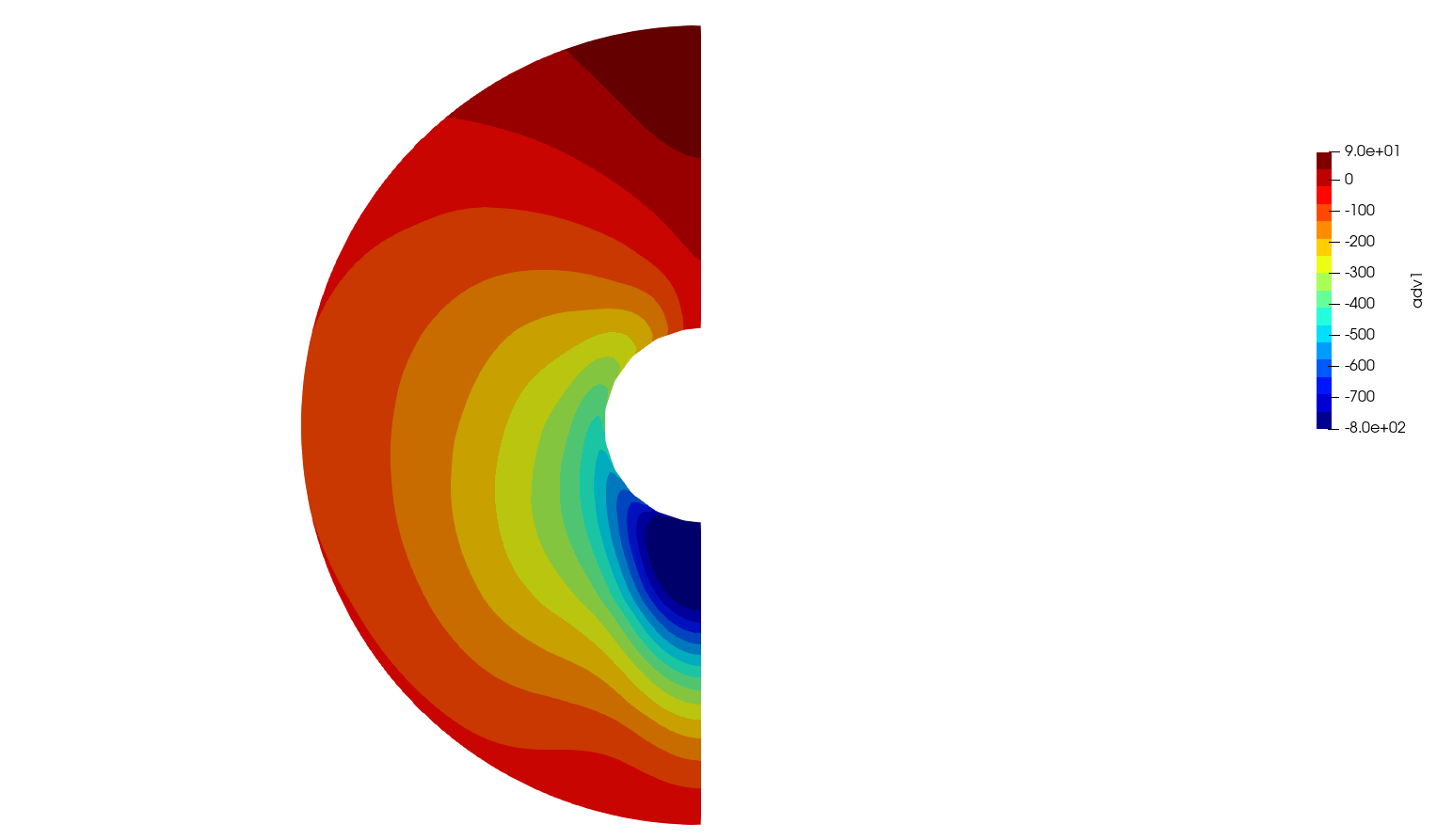}
}
\quad
\subfigure[]{
\includegraphics[height=0.2\textheight, trim=10cm 0cm 28cm 0cm, clip]{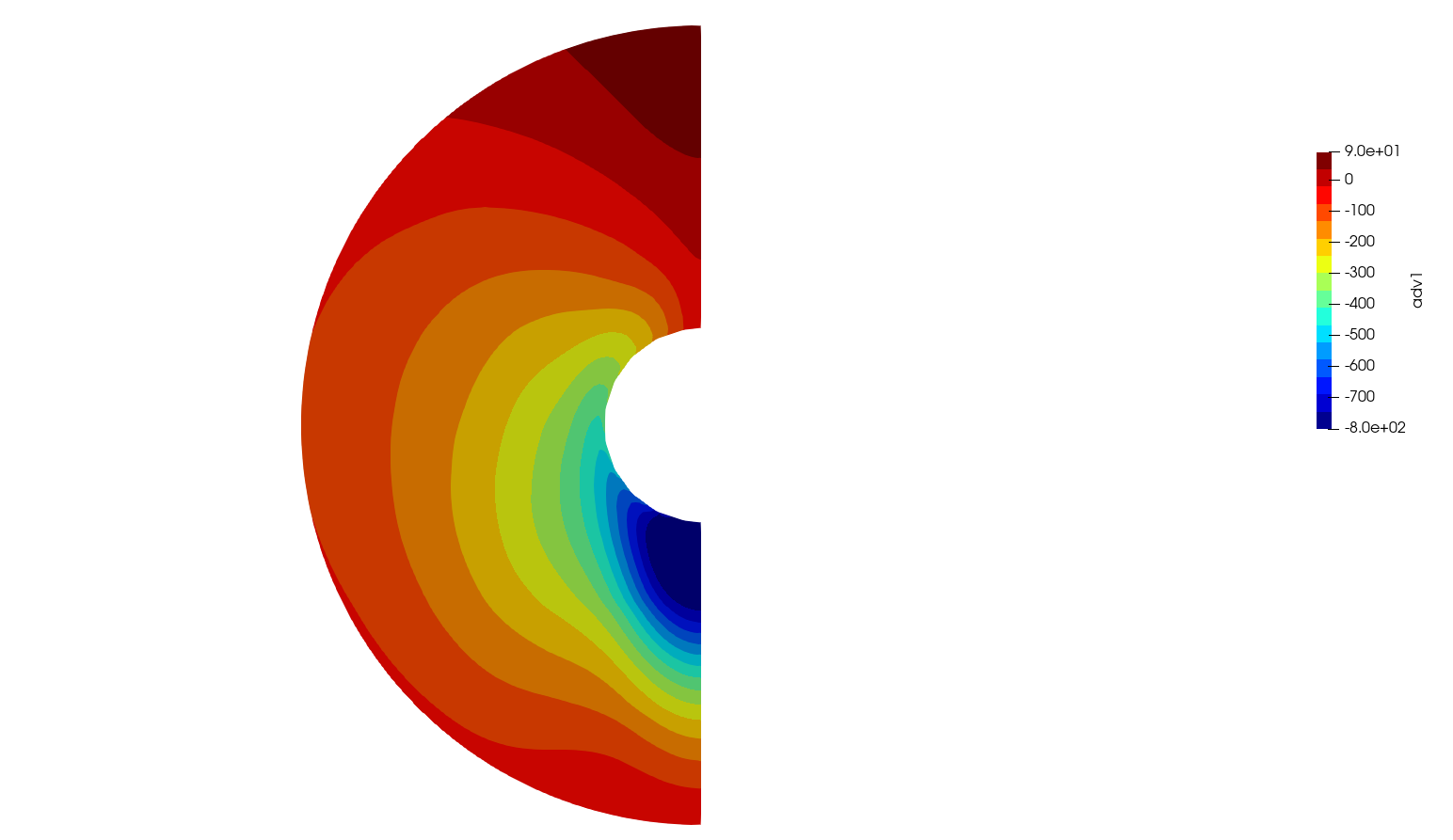}
}
\qquad \qquad
\iftoggle{tikzExternal}{
\input{./tikz/legend_wake_adjoint.tikz}
}{
\includegraphics{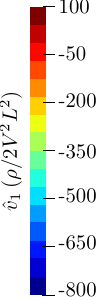}
}
\caption{Japan Bulk Carrier case: Exemplary adjoint axial wake fields. (a)--(b) Classical MWI formulation without correction for $\omega = 0.8$ and $\omega = 0.06$, respectively; (c)--(d) proposed MWI formulation for the same relaxation factors.}
\label{fig:jbc_adjoint_wake}
\end{figure}

Corresponding local flow features are illustrated in Fig.~\ref{fig:jbc_wake}, which shows four exemplary nominal wake fields. Subfigures (a) and (b) correspond to the classical formulation without correction for $\omega = 0.8$ and $\omega = 0.06$, respectively, while (c) and (d) show the corresponding results obtained with the proposed formulation. It is observed that subfigure (b) exhibits a clearly distorted wake structure due to the strong influence of uncorrected relaxation effects. In contrast, the results obtained with the proposed formulation remain consistent and nearly identical across different relaxation factors.
Analogous adjoint axial wake fields are shown in Fig.~\ref{fig:jbc_adjoint_wake}. The overall behavior is consistent with the primal results presented in Fig.~\ref{fig:jbc_wake}. However, the results corresponding to subfigure (b) exhibit a significantly amplified degradation, as expected due to the increased sensitivity of the adjoint solution to solver-induced effects, as already observed in Fig.~\ref{fig:jbc_steady_results_wake}.

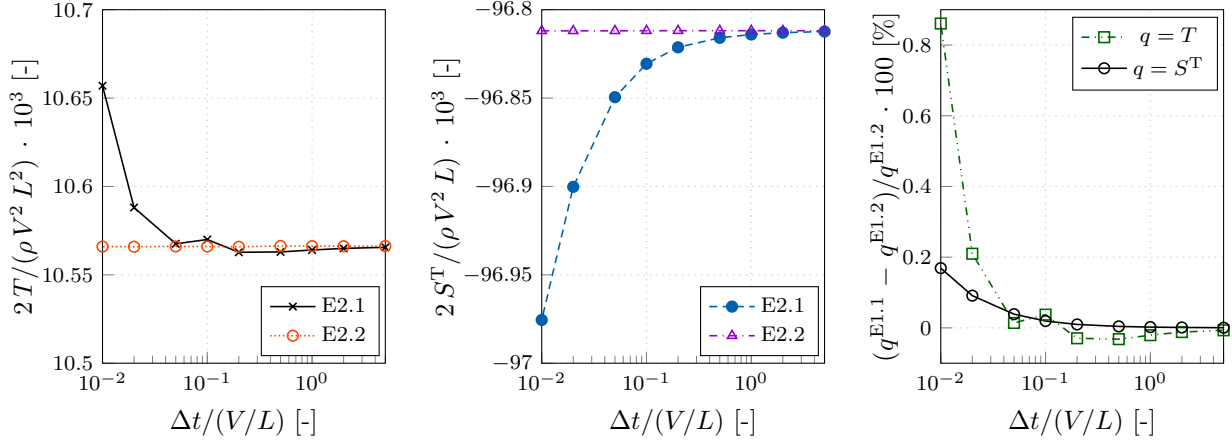
\begin{figure}[!ht]
\centering
\iftoggle{tikzExternal}{
\analytiSolutionPictures
\begin{tikzpicture}
\begin{axis}[
 xlabel={$\Delta t /(V/L)$ [-]},
 xlabel style={text width=0.25\textwidth,align=center},
 ylabel={$2 \, T/(\rho \, V^2 \, L^2) \cdot 10^{3}$ [-]},
 ylabel shift = -2mm,
 ylabel style={text width=0.35\textwidth,align=center},
 legend style={at={(0.98,0.02)},anchor=south east},
 xmode=log,
 xmin=1E-02,
 xmax=5E-00,
 ymin=10.50,
 ymax=10.70,
]

\addplot [line1, mark1, each nth point=1] table[x expr={\thisrowno{0}},y expr={\thisrowno{1}*1E+04}]{data/jbc_results_unsteady.dat};
\addplot [line2, mark2, each nth point=1] table[x expr={\thisrowno{0}},y expr={\thisrowno{2}*1E+04}]{data/jbc_results_unsteady.dat};

\addlegendentry{E2.1};
\addlegendentry{E2.2};

\end{axis}
\end{tikzpicture}
\begin{tikzpicture}
\begin{axis}[
 xlabel={$\Delta t /(V/L)$ [-]},
 xlabel style={text width=0.25\textwidth,align=center},
 ylabel={$2 \, S^\mathrm{T}/(\rho \, V^2 \, L) \cdot 10^{3}$ [-]},
 ylabel shift = -2mm,
 ylabel style={text width=0.35\textwidth,align=center},
 legend style={at={(0.98,0.02)},anchor=south east},
 xmode=log,
 xmin=1E-02,
 xmax=5E-00,
 ymin=-97.0,
 ymax=-96.8,
]

\addplot [line3, mark3, each nth point=1] table[x expr={\thisrowno{0}},y expr={\thisrowno{5}*1E+03}]{data/jbc_results_unsteady.dat};
\addplot [line4, mark4, each nth point=1] table[x expr={\thisrowno{0}},y expr={\thisrowno{6}*1E+03}]{data/jbc_results_unsteady.dat};

\addlegendentry{E2.1};
\addlegendentry{E2.2};

\end{axis}
\end{tikzpicture}
\begin{tikzpicture}
\begin{axis}[
 xlabel={$\Delta t /(V/L)$ [-]},
 xlabel style={text width=0.25\textwidth,align=center},
 ylabel={$(q^\mathrm{E1.1} - q^\mathrm{E1.2})/q^\mathrm{E1.2} \cdot 100$ [\%]},
 ylabel shift = -2mm,
 ylabel style={text width=0.35\textwidth,align=center},
 legend style={at={(0.98,0.98)},anchor=north east},
 xmode=log,
 xmin=1E-02,
 xmax=5E-00,
 ymin=-0.1,
 ymax=0.9,
]

\addplot [line5, mark5, each nth point=1] table[x expr={\thisrowno{0}},y expr={ ((\thisrowno{1}-\thisrowno{2})/\thisrowno{2})*100 }]{data/jbc_results_unsteady.dat};
\addplot [line6, mark6, each nth point=1] table[x expr={\thisrowno{0}},y expr={ ((\thisrowno{5}-\thisrowno{6})/\thisrowno{6})*100 }]{data/jbc_results_unsteady.dat};

\addlegendentry{$q = T$};
\addlegendentry{$q = S^\mathrm{T}$};

\end{axis}
\end{tikzpicture}
}{
\includegraphics{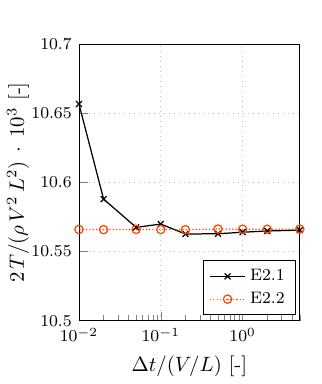}
\includegraphics{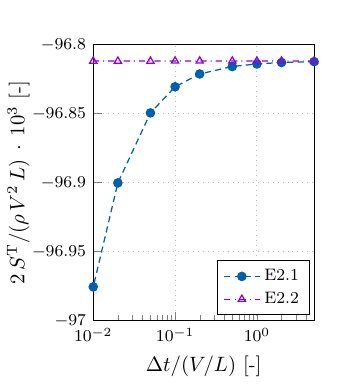}
\includegraphics{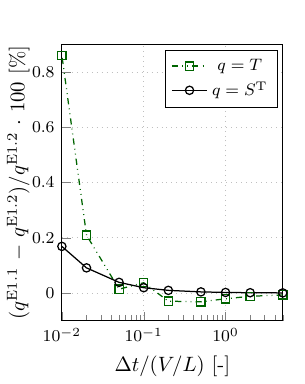}
}
\caption{Japan Bulk Carrier case ($\mathrm{Re} = 10^{7}$): Influence of the normalized time-step size on the propeller thrust $T$ (left) and the corresponding sensitivity $S^\mathrm{T}$ (center) for Experiments E2.1 and E2.2. The right-hand side shows the relative deviation between both formulations.}
\label{fig:jbc_unsteady_results_thrust}
\end{figure}

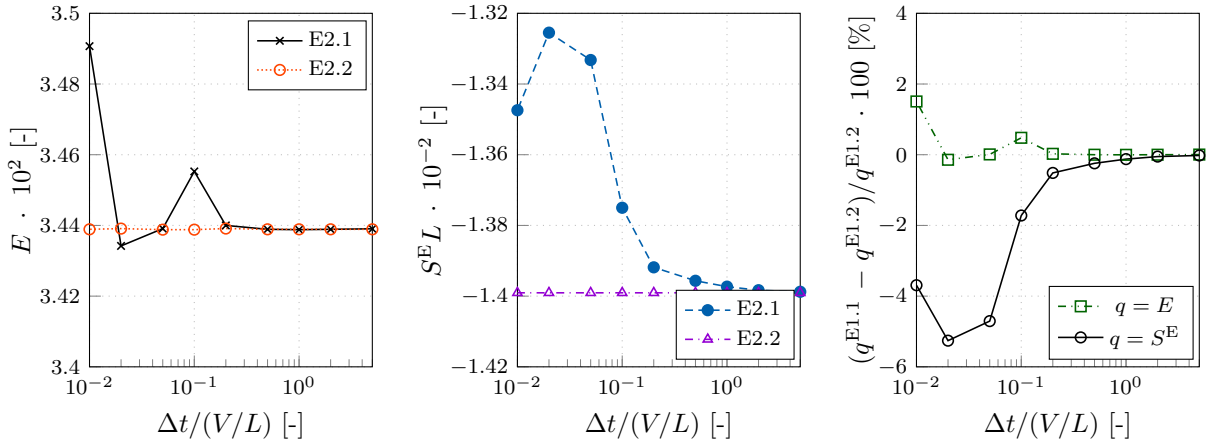
\begin{figure}[!ht]
\centering
\iftoggle{tikzExternal}{
\analytiSolutionPictures
\begin{tikzpicture}
\begin{axis}[
 xlabel={$\Delta t /(V/L)$ [-]},
 xlabel style={text width=0.25\textwidth,align=center},
 ylabel={$E \cdot 10^2$ [-]},
 ylabel shift = -2mm,
 ylabel style={text width=0.35\textwidth,align=center},
 legend style={at={(0.98,0.98)},anchor=north east},
 xmode=log,
 xmin=1E-02,
 xmax=5E-00,
 ymin=3.4,
 ymax=3.5,
]

\addplot [line1, mark1, each nth point=1] table[x expr={\thisrowno{0}},y expr={\thisrowno{3}*1E+02}]{data/jbc_results_unsteady.dat};
\addplot [line2, mark2, each nth point=1] table[x expr={\thisrowno{0}},y expr={\thisrowno{4}*1E+02}]{data/jbc_results_unsteady.dat};

\addlegendentry{E2.1};
\addlegendentry{E2.2};

\end{axis}
\end{tikzpicture}
\begin{tikzpicture}
\begin{axis}[
 xlabel={$\Delta t /(V/L)$ [-]},
 xlabel style={text width=0.25\textwidth,align=center},
 ylabel={$S^\mathrm{E} L \cdot 10^{-2}$ [-]},
 ylabel shift = -2mm,
 ylabel style={text width=0.35\textwidth,align=center},
 legend style={at={(0.98,0.02)},anchor=south east},
 xmode=log,
 xmin=1E-02,
 xmax=5E-00,
 ymin=-1.42,
 ymax=-1.32,
]

\addplot [line3, mark3, each nth point=1] table[x expr={\thisrowno{0}},y expr={\thisrowno{7}*1E-02}]{data/jbc_results_unsteady__coarse.dat};
\addplot [line4, mark4, each nth point=1] table[x expr={\thisrowno{0}},y expr={\thisrowno{8}*1E-02}]{data/jbc_results_unsteady__coarse.dat};

\addlegendentry{E2.1};
\addlegendentry{E2.2};

\end{axis}
\end{tikzpicture}
\begin{tikzpicture}
\begin{axis}[
 xlabel={$\Delta t /(V/L)$ [-]},
 xlabel style={text width=0.25\textwidth,align=center},
 ylabel={$(q^\mathrm{E1.1} - q^\mathrm{E1.2})/q^\mathrm{E1.2} \cdot 100$ [\%]},
 ylabel shift = -2mm,
 ylabel style={text width=0.35\textwidth,align=center},
 legend style={at={(0.98,0.02)},anchor=south east},
 xmode=log,
 xmin=1E-02,
 xmax=5E-00,
 ymin=-6,
 ymax=4,
]

\addplot [line5, mark5, each nth point=1] table[x expr={\thisrowno{0}},y expr={ ((\thisrowno{3}-\thisrowno{4})/\thisrowno{4})*100 }]{data/jbc_results_unsteady.dat};
\addplot [line6, mark6, each nth point=1] table[x expr={\thisrowno{0}},y expr={ ((\thisrowno{7}-\thisrowno{8})/\thisrowno{8})*100 }]{data/jbc_results_unsteady__coarse.dat};

\addlegendentry{$q = E$};
\addlegendentry{$q = S^\mathrm{E}$};

\end{axis}
\end{tikzpicture}
}{
\includegraphics{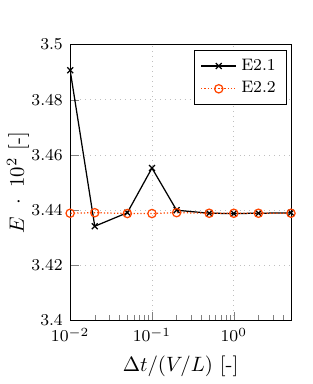}
\includegraphics{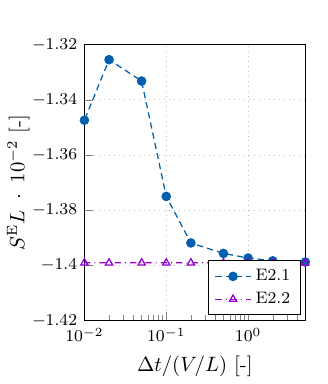}
\includegraphics{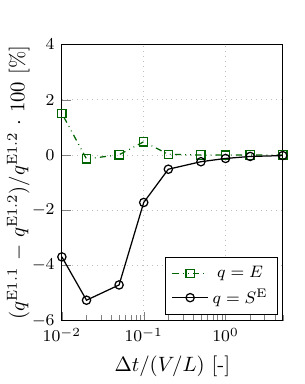}
}
\caption{Japan Bulk Carrier case ($\mathrm{Re} = 10^{7}$): Influence of the normalized time-step size on the effective wake homogeneity $E$ (left) and the corresponding sensitivity $S^\mathrm{E}$ (center) for Experiments E2.1 and E2.2. The right-hand side shows the relative deviation between both formulations.}
\label{fig:jbc_unsteady_results_wake}
\end{figure}

The propulsion case is considered next. The corresponding results are shown in Fig.~\ref{fig:jbc_unsteady_results_thrust}, where the propeller thrust $T$ (left) and the corresponding sensitivity $S^\mathrm{T}$ (center) are plotted over the normalized time-step size. As before, the classical formulation exhibits a clear dependence on the time-step size, leading to varying thrust predictions. In contrast, the proposed formulation (E2.2) yields results that are independent of the time-step size. The relative differences between the two formulations are shown in the right-hand plot and amount to approximately $+1\%$ for thrust and lower values for the corresponding sensitivity.
Subsequently, the wake homogeneity functional under effective conditions, i.e., in propulsion mode, is analyzed. The results for $E$ and $S^\mathrm{E}$ are shown in Fig.~\ref{fig:jbc_unsteady_results_wake}, again plotted over the normalized time-step size. The same trends as observed previously are recovered: neglecting the MWI correction leads to increasing deviations for smaller time-step sizes, whereas the proposed formulation yields consistent results. As expected for self-propulsion conditions, $E$ is smaller than $W$, indicating a more homogeneous flow field; cf.~Fig.~\ref{fig:jbc_steady_results_wake}. For the shape sensitivity, the behavior is again quantitatively similar to the nominal scenario, including a sign change of the integrated sensitivity. The relative differences between the two formulations are shown in the right-hand plot.


\section{Conclusions}
\label{sec:conclusion}
The present work addresses the dependence of the Momentum-Weighted Interpolation (MWI) on solver-specific parameters such as under-relaxation and time-step size. The proposed formulation applies identically to both primal and continuous adjoint flow solvers, ensuring consistency between state and sensitivity computations.

It is shown that the observed dependency arises from the implicit inclusion of solver-induced contributions in the diagonal momentum coefficient that enters the mobility term.

A consistent reformulation of the MWI is proposed in which these contributions are removed from the pressure-driven correction. The derivation follows a unified, generic procedure and is presented for four configurations, including the classical formulation and cases involving relaxation, time discretization, and their combined effect. The formulation is provided in a step-by-step, recipe-like manner, enabling straightforward reproduction and implementation. The proposed modification requires only minimal changes to existing implementations.

The resulting formulation yields a face flux that is independent of both the relaxation factor and the time-step size, while retaining the correct transient and iterative behavior through additional recursive terms.

The approach is assessed for a two-dimensional laminar cylinder flow and a three-dimensional turbulent ship flow. In both cases, the classical formulation exhibits clear dependencies between local and integral flow measures, such as forces and wake characteristics, and solver parameters. In contrast, the proposed formulation removes these dependencies and yields consistent results. 

For the considered integral force-based quantities, deviations between the inconsistent and consistent formulations are on the order of $\mathcal{O}(1\%)$. For integral field-based quantities, significantly larger deviations can occur, reaching up to approximately $20\%$ for the ship wake investigated in this work. Adjoint results exhibit even stronger discrepancies and may change sign.

Overall, the proposed modification enhances the robustness and consistency of pressure--velocity coupling schemes for both primal and adjoint formulations. This is particularly relevant for practical applications, where adjoint systems, although linear, are often more strongly coupled due to the underlying nonlinear primal solution and therefore exhibit increased numerical stiffness.

\section{Acknowledgments}
The current work is part of the “Propulsion Optimization of Ships and Appendages” (Grant No. 03SX599C) research project funded by the German Federal Ministry for Economics and Climate Action. The authors gratefully acknowledge this support.

\section{Acknowledgment of AI Assistance}
AI-based language tools were used to assist in refining the wording and structure of parts of the manuscript. All scientific content, derivations, and conclusions were developed and verified by the authors.

\section{Declaration of Competing Interest}
The authors declare that they have no known competing financial interests or personal relationships that could have appeared to influence the work reported in this paper.

\section{Data Availability Statement}
No new data were generated or analyzed in this study.

\section{CRediT authorship contribution statement}
\textbf{N.~K.}: Conceptualization, Methodology, Software, Validation, Formal analysis, Investigation, Data curation, Visualization, Writing--original draft.
\textbf{D.~H.}: Resources, Writing--review \& editing.



\begin{appendix}

\section{Removal of Relaxation Contributions}
\label{sec:derivation_mwi_relax}
In practical segregated solution procedures, the momentum equations are typically solved in under-relaxed form. In this case, the semi-discrete momentum equation is modified such that
\begin{align}
\frac{A_\mathrm{P}^\mathrm{k}}{\omega} \, v_{\mathrm{P},\mathrm{i}}^\mathrm{k}
=
H_{\mathrm{P},\mathrm{i}}^\mathrm{k}
-
\Omega_\mathrm{P} \left[
\frac{\partial p}{\partial x_{\mathrm{i}}} - f_{\mathrm{i}}
\right]_\mathrm{P}^\mathrm{k}
+
\frac{1-\omega}{\omega} \, A_\mathrm{P}^\mathrm{k} \, v_{\mathrm{P},\mathrm{i}}^\mathrm{k-1},
\label{equ:relax_momentum}
\end{align}
where $\omega \in (0,1]$ denotes the relaxation factor. Rearrangement yields an augmented version of Eqn. \ref{equ:mome_semi_discrete_std}, i.e.
\begin{align}
v_{\mathrm{P},\mathrm{i}}^\mathrm{k}
=
\omega
\left[
\frac{H_{\mathrm{P},\mathrm{i}}}{A_\mathrm{P}^\mathrm{k}}
-
\frac{\Omega_\mathrm{P}}{A_\mathrm{P}^\mathrm{k}}
\left(
\frac{\partial p}{\partial x_{\mathrm{i}}} - f_{\mathrm{i}}
\right)_\mathrm{P}
\right]^\mathrm{k}
+
(1-\omega)\,v_{\mathrm{P},\mathrm{i}}^\mathrm{k-1}.
\end{align}
Proceeding analogously to the standard derivation, the above expression is interpreted both at the cell center and, formally, at the face. Interpolating the cell-centered expression to face $F$ and projecting onto the face area vector yields the predictor flux
\begin{align}
\widetilde{\dot{V}}_\mathrm{F}^\mathrm{k}
&=
\omega
\left(
\frac{H_{\mathrm{P,i}}}{A_\mathrm{P}}
\right)_\mathrm{F}^\mathrm{k} \Delta \Gamma_{\mathrm{F},\mathrm{i}}
-
\omega \, b_\mathrm{F}^\mathrm{k}
\overline{\left[\frac{\partial p}{\partial x_{\mathrm{i}}} - f_{\mathrm{i}}\right]}_\mathrm{F}^\mathrm{k} \Delta \Gamma_{\mathrm{F},\mathrm{i}}
+
(1-\omega)\,\widetilde{\dot{V}}_\mathrm{F}^\mathrm{k-1}.
\label{equ:mwi_predictor_relax}
\end{align}
In analogy to Eqn.~\ref{equ:mwi_face_balance}, a face-based form of the relaxed momentum balance can be written as
\begin{align}
\dot{V}_\mathrm{F}^\mathrm{k}
&=
\omega
\left(
\frac{H_{\mathrm{P,i}}}{A_\mathrm{P}}
\right)_\mathrm{F}^\mathrm{k} \Delta \Gamma_{\mathrm{F},\mathrm{i}}
-
\omega \, b_\mathrm{F}^\mathrm{k}
\left[
\frac{\partial p}{\partial x_{\mathrm{i}}}- f_{\mathrm{i}}
\right]_\mathrm{F}^\mathrm{k}  \Delta \Gamma_{\mathrm{F},\mathrm{i}}
+
(1-\omega)\,\dot{V}_\mathrm{F}^\mathrm{k-1}.
\label{equ:mwi_face_balance_relax}
\end{align}
Equations~\ref{equ:mwi_predictor_relax} and \ref{equ:mwi_face_balance_relax} again contain the same non-pressure-driven contribution. Eliminating this term yields
\begin{align}
\dot{V}_\mathrm{F}^\mathrm{k}
&=
\widetilde{\dot{V}}_\mathrm{F}^\mathrm{k}
-
\omega \, b_\mathrm{F}^\mathrm{k}
\left[
\left[\frac{\partial p}{\partial x_{\mathrm{i}}} - f_{\mathrm{i}}\right]_\mathrm{F}^\mathrm{k}
-
\overline{\left[\frac{\partial p}{\partial x_{\mathrm{i}}} - f_{\mathrm{i}}\right]}_\mathrm{F}^\mathrm{k}
\right]\Delta \Gamma_{\mathrm{F},\mathrm{i}}
+
(1-\omega)
\left(
\dot{V}_\mathrm{F}^\mathrm{k-1}
-
\widetilde{\dot{V}}_\mathrm{F}^\mathrm{k-1}
\right).
\label{equ:mwi_relax_intermediate}
\end{align}
Introducing the stored MWI correction from the previous outer iteration,
\begin{align}
\Delta \dot{V}_\mathrm{F}^\mathrm{k-1}
=
\dot{V}_\mathrm{F}^\mathrm{k-1}
-
\widetilde{\dot{V}}_\mathrm{F}^\mathrm{k-1},
\end{align}
the relaxed MWI formulation can be written in compact recursive form as
\begin{align}
\dot{V}_\mathrm{F}^\mathrm{k}
&=
\widetilde{\dot{V}}_\mathrm{F}^\mathrm{k}
-
\omega \, b_\mathrm{F}^\mathrm{k}
\left[
\left[\frac{\partial p}{\partial x_{\mathrm{i}}} - f_{\mathrm{i}}\right]_\mathrm{F}^\mathrm{k}
-
\overline{\left[\frac{\partial p}{\partial x_{\mathrm{i}}} - f_{\mathrm{i}}\right]}_\mathrm{F}^\mathrm{k}
\right]\Delta \Gamma_{\mathrm{F},\mathrm{i}}
+
(1-\omega)\,\Delta \dot{V}_\mathrm{F}^\mathrm{k-1}.
\label{equ:mwi_relax_final}
\end{align}
%

\section{Removal of Temporal Contributions}
\label{sec:derivation_mwi_time}
In unsteady simulations, the momentum equations contain additional contributions arising from time discretization. Considering implicit Euler time integration, the semi-discrete momentum equation may be written as
\begin{align}
\left(
A_\mathrm{P^\mathrm{n}} + \frac{\rho_\mathrm{P} \Omega_\mathrm{P}}{\Delta t}
\right)v_{\mathrm{P},\mathrm{i}}^\mathrm{n}
=
H_{\mathrm{P},\mathrm{i}}^\mathrm{n}
-
\Omega_\mathrm{P}\left[
\frac{\partial p}{\partial x_{\mathrm{i}}} - f_{\mathrm{i}}
\right]_\mathrm{P}^\mathrm{n}
+
\frac{\rho_\mathrm{P} \Omega_\mathrm{P}}{\Delta t}\,v_{\mathrm{P},\mathrm{i}}^\mathrm{n-1}.
\label{equ:time_momentum}
\end{align}
where $\Delta t$ denotes the time-step size and $\rho_\mathrm{P}$ the density. Dividing Eqn.~\ref{equ:time_momentum} by $A_\mathrm{P}^\mathrm{n}$ yields
\begin{align}
\left(
1 + d_\mathrm{P}^\mathrm{n}
\right)v_{\mathrm{P},\mathrm{i}}^\mathrm{n}
=
\frac{H_{\mathrm{P},\mathrm{i}}^\mathrm{n}}{A_\mathrm{P}^\mathrm{n}}
-
\frac{\Omega_\mathrm{P}}{A_\mathrm{P}^\mathrm{n}}
\left[
\frac{\partial p}{\partial x_{\mathrm{i}}} - f_{\mathrm{i}}
\right]_\mathrm{P}^\mathrm{n}
+
d_\mathrm{P}^\mathrm{n}\,v_{\mathrm{P},\mathrm{i}}^\mathrm{n-1},
\label{equ:time_momentum_scaled}
\end{align}
with the temporal scaling factor
\begin{align}
d_\mathrm{P}^\mathrm{n}
=
\frac{\rho_\mathrm{P} \Omega_\mathrm{P}}{A_\mathrm{P}^\mathrm{n} \Delta t}.
\label{equ:def_dp}
\end{align}
The quantity $d_\mathrm{P}^\mathrm{n}$ represents the cell-centered analogue of the face-based temporal scaling factor introduced below.
Rearrangement yields an augmented version of Eqn.~\ref{equ:mome_semi_discrete_std}, i.e.
\begin{align}
\left(
1 + d_\mathrm{P}^\mathrm{n}
\right)v_{\mathrm{P},\mathrm{i}}^\mathrm{n}
&=
\frac{H_{\mathrm{P},\mathrm{i}}^\mathrm{n}}{A_\mathrm{P}^\mathrm{n}}
-
\frac{\Omega_\mathrm{P}}{A_\mathrm{P}^\mathrm{n}}
\left[
\frac{\partial p}{\partial x_{\mathrm{i}}} - f_{\mathrm{i}}
\right]_\mathrm{P}^\mathrm{n}
+
d_\mathrm{P}^\mathrm{n}\,v_{\mathrm{P},\mathrm{i}}^\mathrm{n-1},
\label{equ:up_time}
\end{align}
The above expression is interpreted both at the cell center and, formally, at the face. Interpolating the cell-centered expression to face $F$ and projecting onto the face area vector yields the predictor flux
\begin{align}
\left(
1 + d_\mathrm{F}^\mathrm{n}
\right)\widetilde{\dot{V}}_\mathrm{F}^\mathrm{n}
&=
\left(
\frac{H_{\mathrm{P,i}}}{A_\mathrm{P}^\mathrm{n}}
\right)_\mathrm{F}^\mathrm{n} \Delta \Gamma_{\mathrm{F},\mathrm{i}}
-
b_\mathrm{F}^\mathrm{n}
\overline{\left[\frac{\partial p}{\partial x_{\mathrm{i}}} - f_{\mathrm{i}}\right]}_\mathrm{F}^\mathrm{n} \Delta \Gamma_{\mathrm{F},\mathrm{i}}
+
d_\mathrm{F}^\mathrm{n}\,\widetilde{\dot{V}}_\mathrm{F}^\mathrm{n-1},
\label{equ:mwi_predictor_time}
\end{align}
where $d_\mathrm{F}^\mathrm{n}$ denotes the face-based temporal scaling factor obtained from interpolation of $d_\mathrm{P}^\mathrm{n}$ to the face.
In analogy to Eqn.~\ref{equ:mwi_face_balance}, a face-based form of Eqn.~\ref{equ:up_time} can be written as
\begin{align}
\left(
1 + d_\mathrm{F}^\mathrm{n}
\right)\dot{V}_\mathrm{F}^\mathrm{n}
&=
\left(
\frac{H_{\mathrm{P,i}}}{A_\mathrm{P}^\mathrm{n}}
\right)_\mathrm{F}^\mathrm{n} \Delta \Gamma_{\mathrm{F},\mathrm{i}}
-
b_\mathrm{F}^\mathrm{n}
\left[
\frac{\partial p}{\partial x_{\mathrm{i}}} - f_{\mathrm{i}}
\right]_\mathrm{F}^\mathrm{n} \Delta \Gamma_{\mathrm{F},\mathrm{i}}
+
d_\mathrm{F}^\mathrm{n}\,\dot{V}_\mathrm{F}^\mathrm{n-1}.
\label{equ:mwi_face_balance_time}
\end{align}
Equations~\ref{equ:mwi_predictor_time} and \ref{equ:mwi_face_balance_time} again contain the same non-pressure-driven contribution. Eliminating this term yields
\begin{align}
\left(
1 + d_\mathrm{F}^\mathrm{n}
\right)\dot{V}_\mathrm{F}^\mathrm{n}
&=
\left(
1 + d_\mathrm{F}^\mathrm{n}
\right)\widetilde{\dot{V}}_\mathrm{F}^\mathrm{n}
-
b_\mathrm{F}^\mathrm{n}
\left[
\left[\frac{\partial p}{\partial x_{\mathrm{i}}} - f_{\mathrm{i}}\right]_\mathrm{F}^\mathrm{n}
-
\overline{\left[\frac{\partial p}{\partial x_{\mathrm{i}}} - f_{\mathrm{i}}\right]}_\mathrm{F}^\mathrm{n}
\right]\Delta \Gamma_{\mathrm{F},\mathrm{i}}
\nonumber\\
&\quad
+
d_\mathrm{F}^\mathrm{n}
\left(
\dot{V}_\mathrm{F}^\mathrm{n-1}
-
\widetilde{\dot{V}}_\mathrm{F}^\mathrm{n-1}
\right).
\label{equ:mwi_time_intermediate}
\end{align}
Introducing the stored MWI correction from the previous time level,
\begin{align}
\Delta \dot{V}_\mathrm{F}^\mathrm{n-1}
=
\dot{V}_\mathrm{F}^\mathrm{n-1}
-
\widetilde{\dot{V}}_\mathrm{F}^\mathrm{n-1},
\end{align}
the time-dependent MWI formulation can be written in compact recursive form as
\begin{align}
\dot{V}_\mathrm{F}^\mathrm{n}
&=
\widetilde{\dot{V}}_\mathrm{F}^\mathrm{n}
-
\frac{b_\mathrm{F}^\mathrm{n}}{1 + d_\mathrm{F}^\mathrm{n}}
\left[
\left[\frac{\partial p}{\partial x_{\mathrm{i}}} - f_{\mathrm{i}}\right]_\mathrm{F}^\mathrm{n}
-
\overline{\left[\frac{\partial p}{\partial x_{\mathrm{i}}} - f_{\mathrm{i}}\right]}_\mathrm{F}^\mathrm{n}
\right]\Delta \Gamma_{\mathrm{F},\mathrm{i}}
+
\frac{d_\mathrm{F}^\mathrm{n}}{1 + d_\mathrm{F}^\mathrm{n}}\,\Delta \dot{V}_\mathrm{F}^\mathrm{n-1}.
\label{equ:mwi_time_final}
\end{align}
%

\end{appendix}

\end{document}